\documentclass[aps,prd,preprint]{revtex4}
\usepackage{amsmath}
\usepackage{amsfonts}
\usepackage{amssymb}
\usepackage{bm}
\usepackage{enumerate}
\usepackage[dvips]{color,graphicx}
\usepackage{epsfig}
\usepackage[figuresright]{rotating}

\newcommand{\Fntwiddle} {\widetilde{{\cal F}}^n(x)}

\newcommand{\xonyQ} {\left ( \frac{x}{y},Q^2 \right )}

\begin{document}

\title{New Method for Extracting Neutron Structure Functions \\
	from Nuclear Data}

\author{Yonatan Kahn}
\affiliation{Northwestern University, Evanston, Illinois 60208, USA, and \\
         Jefferson Lab, Newport News, Virginia 23606, USA}
\author{W. Melnitchouk}
\affiliation{Jefferson Lab, Newport News, Virginia 23606, USA}
\author{S. A. Kulagin}
\affiliation{Institute for Nuclear Research, Moscow 117312, Russia\\ }

\begin{abstract}
We propose a new method for extracting neutron structure functions
from inclusive structure functions of nuclei, which employs an 
iterative procedure of solving integral convolution equations.
Unlike earlier approaches, the new method is applicable to both
spin-averaged and spin-dependent structure functions.
We test the reliability of the method on unpolarized $F_2$ and
polarized $g_1$ structure functions of the deuteron in both the
nucleon resonance and deep inelastic regions.
The new method is able to reproduce known input functions of almost
arbitrary shape to very good accuracy with only several iterations.
\end{abstract}

\maketitle

\section{Introduction}

Understanding the detailed structure of the nucleon remains one of
the central problems of the strong nuclear interactions.
This is particularly challenging in the so-called transition region
at momentum scales $\sim 1$~GeV, where neither perturbative Quantum
Chromodynamics (QCD) nor effective hadronic theories provide
adequate descriptions of physical observables.

Over the past few years one of the fascinating developments in the 
study of this transition has been the phenomenon of quark-hadron 
duality in inclusive electron--nucleon (and nucleus) scattering.
Here the structure functions in the region dominated by low-lying
resonant excitations of the nucleon are found to closely resemble,
on average, the deep inelastic structure functions describing the
high-energy cross section \cite{BG}.
Since QCD at high energy and momentum transfers can be treated 
perturbatively, but is highly nonperturbative at low energies where 
hadronic degrees of freedom are prominent, this duality provides an 
intimate link between the two regimes.

Recent experiments have sought to quantify quark-hadron duality by
determining its flavor, spin, and nuclear dependence, while theoretical
endeavors have attempted to understand its dynamical origin from a more
fundamental basis (for a review see Ref.~\cite{MEK}).
In particular, even from simple quark model arguments one expects
intriguingly different behaviors of duality for the proton and for the
neutron \cite{Theory}.

While duality for the proton has been tested to rather good accuracy
in recent measurements, for both unpolarized and polarized scattering
\cite{Exp_p,Ales}, there is almost a complete absence of analogous
empirical information on the neutron.
This lack of knowledge has prevented the various theoretical models
from being adequately tested, and has impeded progress in unraveling
the microscopic origin of the duality phenomenon.
The difficulty with obtaining data on neutron structure functions
is of course the absence of free neutron targets.
As a result one often makes use of light nuclei such as deuterium
\cite{Osipenko} or $^3$He \cite{Exp_He3} as effective neutron targets,
assuming that the nuclear corrections are negligible.

Even when nuclear effects are considered, there exist practical
difficulties with extracting information on the neutron from nuclear
data.
Some attempts have been made to obtain the spin-averaged $F_2^n$
structure function from proton and deuterium data in the deep inelastic
scattering region, where the exchanged four-momentum transfer squared
$Q^2$ is large ($\sim~5$~GeV$^2$ or greater).
A common approach has made use of the so-called smearing factor method
\cite{Bodek}, where after an initial guess for $F_2^n$ one iterates
the solution in order to eliminate the dependence of the extracted
neutron structure function on the starting point.

In practice the smearing factor method has only been applied to the
unpolarized $F_2$ structure function and only in deep inelastic
kinematics \cite{Whitlow}.
The robustness of this procedure is guaranteed only for functions
which do not change sign, and for spin-dependent structure functions,
which can have several zeros, the usual prescription is inadequate.
Furthermore, in the nucleon resonance region, where there exists
non-trivial resonant structure, it is not {\em a priori} clear
whether it is even possible to extract resonance structure that has
been smeared out by nucleon Fermi motion.

In this paper we propose a new method in which the nuclear effects are
parameterized via an additive correction to the free nucleon structure
functions.
In contrast to the more common multiplicative method, which is
problematic for structure functions with zeros, the new method can
be used for functions of almost arbitrary shape, which allows access
to neutron structure in both the deep inelastic and resonance regions.
By iterating the solution, the dependence on the initial guess for the
neutron structure function is eliminated, and in practice a reliable
extraction can be achieved after only several iterations.

In Sec.~\ref{sec:nuclearSF} we present the formalism for computing
nuclear structure functions at finite $Q^2$ within the nuclear impulse
approximation.
While the formalism is general and can be applied to any nucleus,
to illustrate the features of the new extraction method we focus on
the specific case of the deuteron.
In Sec.~\ref{sec:methods} we present the details of the new method,
and discuss other methods which have been used to extract neutron
structure functions from nuclear data, including the smearing factor
method, and a direct method of inverting integral equations which has
previously been used in Refs.~\cite{Umnikov,Scopetta}.
Our results are presented in Sec.~\ref{sec:results} for spin-averaged
and spin-dependent structure functions, in both the resonance and
deep inelastic regions.
Using known input functions constructed from resonance and leading
twist structure function parameterizations, we demonstrate the
accuracy of the extraction method and provide a detailed discussion
of its convergence to the exact results as a function of the number
of iterations and the first guess in the iteration.
Finally, in Sec.~\ref{sec:conclusion} we summarize our results
and preview future applications of the new method.

\section{Nuclear structure functions}
\label{sec:nuclearSF}

The usual framework for computing structure functions of nuclei at large
$x$ is the relativistic nuclear impulse approximation, in which the
lepton probe scatters from the nucleus incoherently via the scattering
from its bound proton and neutron constituents.
In this approximation the nuclear structure functions can be written
as convolutions of the bound nucleon structure functions and nucleon
light-cone momentum distributions in the nucleus
\cite{MST,KPW,MPT,KMPW,KP,AKL,KM_d,KM_He3}.

In particular, for the spin-averaged $F_2$ structure function of a
nucleus $A$ we have:
\begin{eqnarray}
F_2^A(x,Q^2)
&=& \left( f_0^{p/A} \otimes F_2^p \right)(x,Q^2)\
 +\ \left( f_0^{n/A} \otimes F_2^n \right)(x,Q^2)\ ,
\label{eq:conv_F}
\end{eqnarray}
where $x = Q^2/2M_A\nu$ is the Bjorken scaling variable (per nucleon),
$M_A$ is the nuclear mass and $\nu$ is the energy transfer, and the
symbol $\otimes$ denotes the convolution
\begin{eqnarray}
\left( f_0^{N/A} \otimes F_2^N \right)(x,Q^2)
&\equiv& \int_x^{M_A/M} dy\ f_0^{N/A}(y,\gamma)\ F_2^N\xonyQ\ ,
\label{eq:conv_def}
\end{eqnarray}
with $M$ the nucleon mass.
The function $f_0^{N/A}$ is the light-cone momentum distribution of
nucleons $N$ in the nucleus, and is a function of the light-cone
momentum fraction $y$ of the nucleus carried by protons ($N=p$)
or neutrons ($N=n$), and of the virtual photon ``velocity''
$\gamma$ in the target rest frame,
$\gamma = |{\bm q}|/q_0 = (1 + 4 M^2 x^2/Q^2)^{1/2}$.
For moderate $Q^2$ values $Q^2 \sim 1-10$~GeV$^2$, $\gamma$ ranges
between unity and $\approx 2$.
As discussed in Ref.~\cite{KM_d,KM_He3}, taking the full $Q^2$
dependence of the smearing function into account is vital for
discussing nuclear structure functions at large-$x$ or resonance
kinematics.

For the spin-dependent nuclear $g_1^A$ and $g_2^A$ structure functions,
one has a set of coupled equations involving both the nucleon $g_1^N$
and $g_2^N$ structure functions \cite{KM_d},
\begin{eqnarray}
xg_i^A(x,Q^2)
&=& \left( f_{ij}^{p/A} \otimes xg_j^p \right)(x,Q^2)\
 +\ \left( f_{ij}^{n/A} \otimes xg_j^n \right)(x,Q^2)\ ,\ \ \ \
    i,j=1,2
\label{eq:conv_g}
\end{eqnarray}
where $f_{ij}^{N/A}$ are the spin-dependent nucleon light-cone momentum
distribution functions in the nucleus, and a sum over repeated indices
$j$ is implied.
In contrast to $F_2^A$, which receives contributions only from the
nucleon $F_2^N$ structure function, the spin-dependent structure
functions at finite $Q^2$ involve also non-diagonal contributions
$f_{12}^{N/A}$ and $f_{21}^{N/A}$.
(Note that both the transverse $F_T^A$ and longitudinal $F_L^A$
structure functions individually receive non-diagonal contributions,
whereas $F_2^A$ does not \cite{KP}.)
In the Bjorken limit, the distribution $f_{12}^{N/A}$ vanishes, and
the expression for the $g_1$ structure function becomes diagonal.
Equations~(\ref{eq:conv_F}) and (\ref{eq:conv_g}) can be viewed as
equations in the single independent variable $x$ for fixed values
of $Q^2$.
In the following, for ease of notation we suppress the dependence
of the structure functions on $Q^2$.

The light-cone momentum distribution functions in Eqs.~(\ref{eq:conv_F}) 
and (\ref{eq:conv_g}) (also referred to as \emph{smearing functions})
can in general be calculated from nuclear spectral functions which
account for the ground state wave function of the nucleus and the
excitation spectrum of the spectator nuclear system, including the
continuum spectrum.
Since the characteristic energies and momenta of the bound nucleons are
small compared with the nucleon mass $M$, the unpolarized distribution
$f_0$ and the spin-dependent diagonal distributions $f_{11}$ and
$f_{22}$ are sharply peaked about $y=1$.


In this analysis we will focus on the case of the deuteron, for which
the smearing functions have recently been evaluated in the weak binding
approximation \cite{KP,KM_d}, including the finite-$Q^2$ corrections
encoded through the dependence on $\gamma$.
%
%
Note that in the isospin symmetric limit the proton and neutron
distributions in the deuteron are identical,
$f^{p/d} = f^{n/d} \equiv f$, and we shall in the following omit the
superscripts on these distributions.

The unpolarized $f_0$ distribution function is given in terms
of the deuteron wave function $\psi_d(p)$ by \cite{KP,AKL}
\begin{eqnarray}
f_0(y,\gamma)
&=& \int {{\rm d}^3p \over (2\pi)^3}
   \left| \psi_d(p) \right|^2
   \left( 1 + \frac{\gamma p_z}{M} \right)
   {1 \over \gamma^2}
   \left[
     1 + {(\gamma^2-1) \over y^2}
	 \left( 1 + {2 \varepsilon \over M}
		  + {\bm{p}^2 \over 2M^2} (1-3\widehat{p}_z^2)
	 \right)
   \right]		\nonumber\\
& & \hspace*{2cm}
\times\ \delta\left( y-1-\frac{\varepsilon+\gamma p_z}{M} \right)\ ,
\label{eq:f0}
\end{eqnarray}
where $\varepsilon = \varepsilon_d - \bm p^2/(2M)$, with
$\varepsilon_d = -2.2$~MeV the deuteron binding energy.
The analogous spin-dependent light-cone distributions $f_{ij}$ are
given explicitly in Ref.~\cite{KM_d}.
For $\gamma=1$ the $f_0$ distribution is normalized to the number of
protons or neutrons in the deuteron, while $f_{11}$ is normalized
to the nucleon polarization in the deuteron,
\begin{eqnarray}
\int_0^{M_A/M} dy\ f_0(y,1)
&=& 1\ ,                \\
\int_0^{M_A/M} dy\ f_{11}(y,1) 
&=& 1 - {3 \over 2}\ \omega_d\ ,
\end{eqnarray}
where $\omega_d$ is the deuteron $D$-state probability.
For the Paris deuteron wave function \cite{Paris} used here
$\omega_d = 5.8\%$.
At finite $Q^2$, or $\gamma > 1$, these normalization conditions are
no longer satisfied, and the distributions do not have probabilistic
interpretations.

\begin{figure}[t] 
\includegraphics[width=10.5cm]{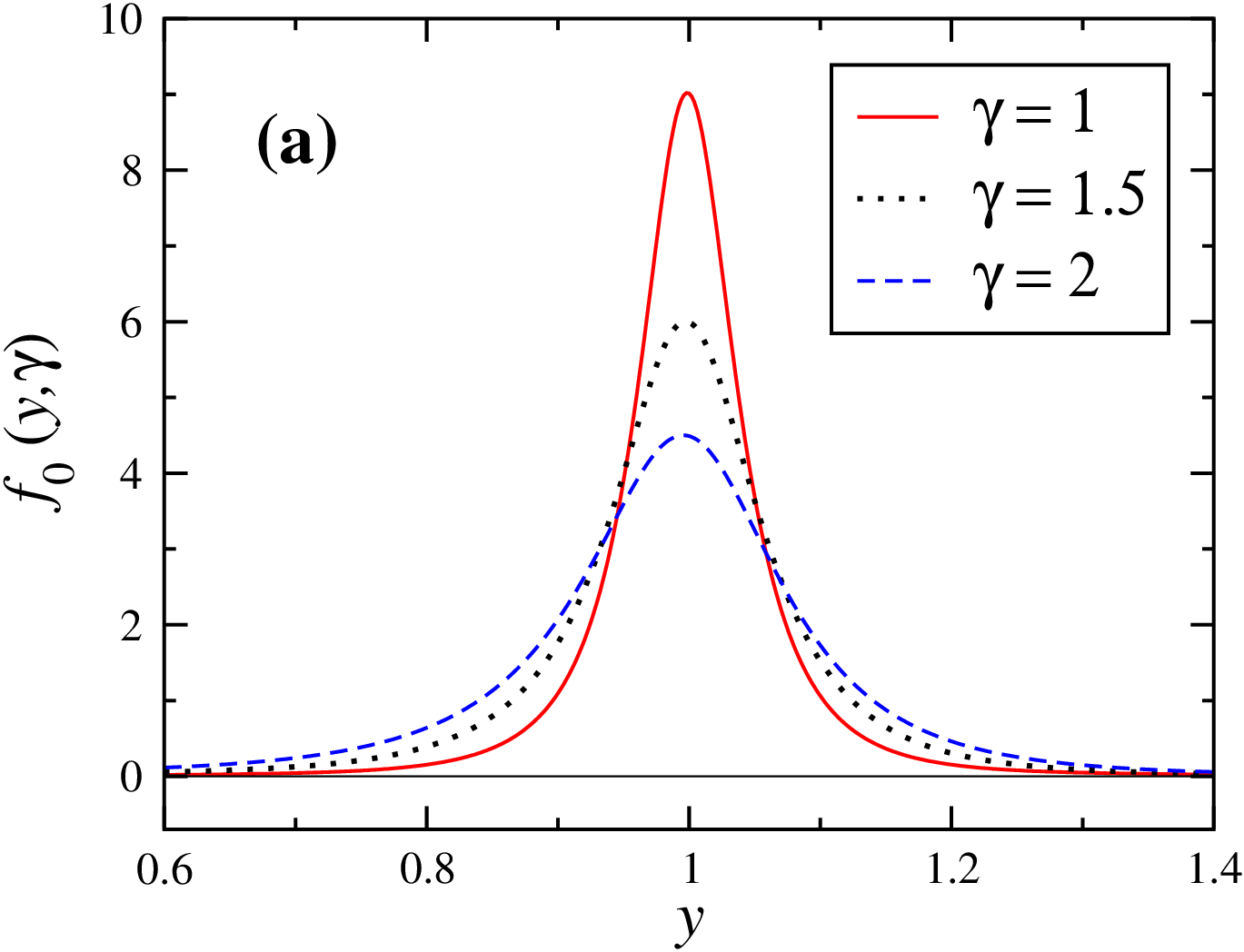}\vspace*{1.2cm}
\includegraphics[width=10.5cm]{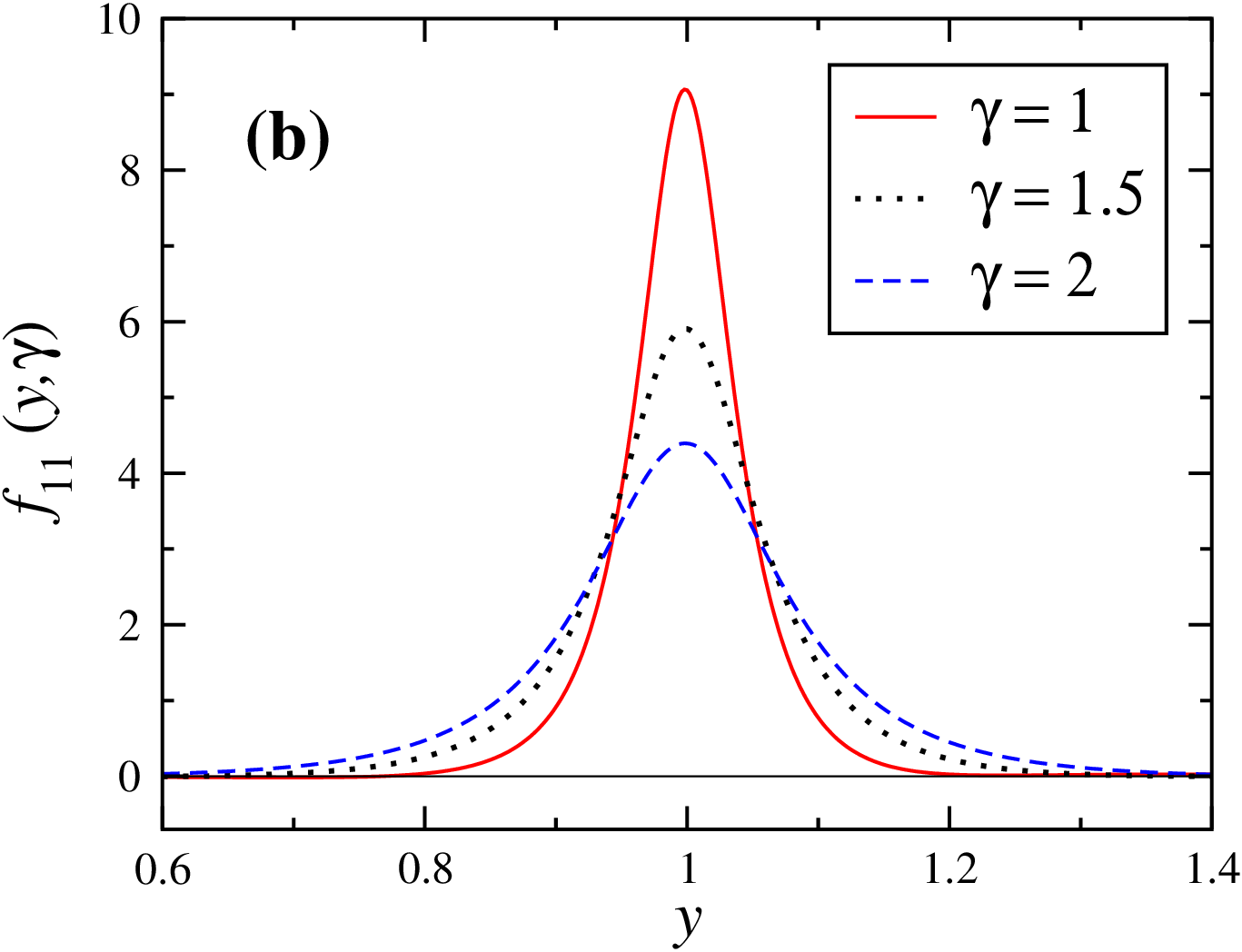}
\caption{(Color online)
    Nucleon distribution functions in the deuteron for
    $\gamma=1$ (solid), 1.5 (dotted) and 2 (dashed):
    (a) $f_0(y,\gamma)$ distribution for the $F_2^d$ structure
    function, (b) $f_{11}(y,\gamma)$ distribution for the
    $xg_1^d$ structure function.}
\label{fig:fy}
\end{figure}

In Fig.~\ref{fig:fy} we show the $f_0(y,\gamma)$ distribution for the
unpolarized $F_2^d$ structure function \cite{KP} and the diagonal
$f_{11}(y,\gamma)$ distribution for the polarized $xg_1^d$ structure
function \cite{KM_d}, for $\gamma=1$, 1.5 and 2, using the deuteron
wave function obtained from the Paris nucleon--nucleon potential
\cite{Paris}.
For $\gamma=1$ the ($Q^2$-independent) distributions are peaked 
sharply around $y=1$, and decrease rapidly with increasing $|y-1|$,
so that by $|y-1| > 0.4$ they become almost negligible.
For larger $\gamma$ the distributions become broader, with a larger
width and smaller height at the peak.
For $\gamma=2$ the height of the peak is approximately half of that
for $\gamma=1$.

The shapes of the unpolarized and polarized distributions are very
similar, with $\approx 0.5\%$ differences between $f_0$ and $f_{11}$
at the peak for $\gamma=1$, and $\lesssim 2.5\%$ for $\gamma=2$.
While the spin-averaged function $f_0$ is constrained to be positive,
the spin-dependent $f_{11}$ function, which involves a difference of
distributions of nucleons with spins aligned and anti-aligned with
that of the deuteron, need not be positive.
For values of $y < 0.8$ the $f_{11}(y,\gamma=1)$ distribution in fact
becomes slightly negative, as is (barely) visible in Fig.~\ref{fig:fy},
although the smearing functions here are close to zero.

Before proceeding to the discussion of extraction methods using these
distributions, we should note that while the impulse approximation
(scattering from individual nucleons in the nucleus) provides the
main contribution to nuclear deep inelastic scattering, in realistic
calculations of nuclear structure functions other effects are also
known to play a role.
These include nuclear shadowing and meson exchange currents at small $x$,
final state interactions of the produced hadronic state, relativistic
effects, and off-shell corrections to the bound nucleon structure
functions \cite{KP}.
Some of these may be formulated within generalized convolutions,
either as two-dimensional convolutions with off-shell structure
functions \cite{MST,KPW,KMPW}, or in terms of exchanged-meson smearing
functions \cite{MEC}, and the techniques discussed here may be
applicable.
Others, such as relativistic corrections, go beyond the convolution
approximation \cite{MST}, and must be included as additive corrections
to the convolution.
Explicit calculations of final state interactions in the quasi-elastic
region have suggested that rescattering effects decrease with increasing
$Q^2$ \cite{Golak}, and in addition partly cancel in inclusive inelastic
cross sections when summed over several exclusive channels \cite{HLM80}.

In the present analysis we do not attempt to provide a complete
description of nuclear structure functions; instead we wish to study
the usefulness of the new method of unsmearing nucleon structure
functions within the conventional convolution framework.
Once we establish the methodology of the new method, additional
effects beyond the convolution approximation can be considered in
actual data analyses.

\section{Extraction methods}
\label{sec:methods}

Having outlined the formalism for computing structure functions of
nuclei in terms of those of nucleons, in this section we review
several methods for extracting neutron structure functions from
nuclear (in practice, deuterium) data, including the new ``additive''
method proposed in this paper.
To extract the neutron structure function from proton and nuclear data
at a given $Q^2$, one first convolutes (or \emph{smears}) the proton
structure function ${\cal F}^p$, where ${\cal F} = F_2$ or $xg_{1,2}$,
with the appropriate smearing function,
\begin{eqnarray}
\widetilde{{\cal F}}^p(x)
&\equiv& \left( f \otimes {\cal F}^p \right)(x)\ ,
\label{eq:p_smear}
\end{eqnarray}
where $f = f_0$ for the unpolarized $F_2$ structure function,
and $f = f_{ij}$ for the polarized $g_{1,2}$ structure functions.
Subtracting the smeared proton $\widetilde{{\cal F}}^p$ from the
nuclear structure function, one obtains an effective smeared neutron
structure function
\begin{eqnarray}
\widetilde{{\cal F}}^n(x)
&=& {\cal F}^d(x) - \widetilde{{\cal F}}^p(x)\ ,
\end{eqnarray}
and then solves the equation
\begin{equation}
\widetilde{{\cal F}}^n(x)
= \left( f \otimes {\cal F}^n \right)(x)
\label{eq:Nconv}
\end{equation}
for ${\cal F}^n(x)$.
Note that for a fixed $Q^2$, $\gamma$ is a function of $x$ alone,
so in practice the smearing functions acquire an $x$ dependence.

\subsection{Direct solution}
\label{ssec:direct}

Equation~(\ref{eq:Nconv}) is a system of so-called Volterra integral
equations of the first kind, which take the general form
\begin{equation}
g(x) = \int_x^{y_{\rm max}} dy\ {K(x,y)\ z(y)}\ ,
\label{Volt}
\end{equation}
where $g(x)$ and $K(x,y)$ (the kernel) are known functions and
$z$ is unknown.
The general theory of Volterra equations is quite extensive,
see for example Ref.~\cite{Volterra}.
Most Volterra equations have no closed-form solution, but numerical
solutions for first-kind equations are quite simple.
Dividing the interval $0 < y < y_{\rm max}$ into a grid of width $h$
by $y_a = ah$, with $a = 0,1,\ldots,N$, and using a quadrature method
such as the trapezoidal rule or Simpson's rule, one can approximate 
the integral in Eq.~(\ref{Volt}) by a discrete sum
\begin{equation}
g_a = \sum_{b=a}^{N} K_{ab}\ z_b\ ,
\label{disc}
\end{equation}
reducing the numerical solution to a problem of matrix inversion:
$\mathbf{z} = \mathbf{K}^{-1}\mathbf{g}$.
In fact, because of the variable lower limit of integration $y=x$,
the matrix $\mathbf{K}$ is upper-triangular, and the inversion is
almost trivial.
This method has been utilized in Refs.~\cite{Umnikov,Scopetta}
in a similar application.
The method fails, though, if $\mathbf{K}$ is singular.

Letting $t = x/y$ and $v = x/y_{\rm max}$, Eq.~(\ref{eq:Nconv})
can be expressed in the form of Eq.~(\ref{Volt}):
\begin{equation}
\widetilde{{\cal F}}^n(y_{\rm max}v)
= \int_v^1 dt\
  f \left( \frac{y_{\rm max}v}{t},\gamma \right)\,
  \frac{y_{\rm max}v}{t^2}\, {\cal F}^n(t)\ ,
\label{NVolt}
\end{equation}
in which case the kernel is a sum of terms
$K(v,t) = (y_{\rm max}v/t^2)\ f(y_{\rm max}v/t,\gamma)$.
The diagonal $K(v,v) \propto f(y_{\rm max})$ corresponds to the
diagonal elements $K_{bb}$ in the discretized equation (\ref{disc}).
However, for any value of $\gamma$, $f(y_{\rm max})$ is extremely
small for strong physical reasons: a single nucleon has a vanishing
probability of carrying the entire momentum of the nucleus.
Thus the matrix $\mathbf{K}$ has very small values along the diagonal
and is very close to singular, so this solution method fails.

A standard approach to solving Volterra equations with a kernel 
vanishing identically along the diagonal is to either integrate 
Eq.~(\ref{NVolt}) by parts, or to differentiate with respect to $x$.
The first technique gives an integral equation for the primitive of
${\cal F}^n(x)$ with kernel $\partial K(x,t)/\partial t$, while the
second has kernel $\partial K(x,t)/\partial x$ and left-hand-side
$d\widetilde{{\cal F}}^{n}(x)/dx$.
These approaches are still problematic, however, because derivatives
of the smearing functions are still very small at $y = y_{\rm max}$;
also, taking derivatives of functions derived from fits to data
introduces substantial errors.
Furthermore, the solution depends on knowing $\Fntwiddle$ at \emph{all}
values of $x$, while in practice, only data up to $x = 1$ are available.
It is clear that a direct solution to Eq.~(\ref{eq:Nconv}) is
impractical for the particular forms of smearing functions used
in this model.

\subsection{Multiplicative solution}
\label{ssec:mult}

The most widely-used method for extracting spin-averaged structure
functions is the smearing-factor or \emph{multiplicative method}
\cite{Bodek}.
This is an iterative solution method based on the {\em ansatz} that
the right-hand-side of Eq.~(\ref{eq:Nconv}) can be written as a product
of the neutron structure function and a ``smearing factor'' $S^n(x)$,
\begin{equation}
\widetilde{\cal F}^n(x)\
=\ S^n(x)\ {\cal F}^n(x)\ .
\label{eq:mult}
\end{equation}
From a first guess ${\cal F}^{n(0)}(x)$, one obtains $S^{(0)}(x)$ by
smearing ${\cal F}^{n(0)}(x)$ and dividing by ${\cal F}^{n(0)}(x)$.
Dividing $\widetilde{\cal F}^n(x)$ by $S^{n(0)}$ gives
${\cal F}^{n(1)}(x)$, so the result after one iteration is
\begin{equation}
{\cal F}^{n(1)}(x)\
=\ \widetilde{\cal F}^n(x)
  { {\cal F}^{n(0)}(x) \over
    \left( f \otimes {\cal F}^{n(0)} \right)(x)
  }\ .
\label{mult1it}
\end{equation}
One can see from the form of Eq.~(\ref{mult1it}) that this method is
problematic if the smeared structure function has zeros in the range
of $x$ of interest.
The spin-averaged nuclear structure functions are positive-definite
for $0 < x < 1$, so this problem does not arise, and the multiplicative
method converges quite rapidly for essentially any reasonable choice of
${\cal F}^{n(0)}(x)$.
Even for spin-dependent structure functions, which may have several
zeros, the multiplicative method works fine as long as the zeros of
the smeared ${\cal F}^{n(0)}(x)$ are very close to the zeros of
$\widetilde{\cal F}^n(x)$.
Since the smearing functions are close to $\delta$-functions, this
amounts to requiring that the zeros of the neutron structure function
be very close to the zeros of the nuclear structure function.
Experimental errors could easily obscure the true location of the zeros
of the nuclear structure function, though, making a direct application
of this method to experimental data difficult.

\subsection{Additive extraction method}
\label{ssec:add}

Instead of assuming a multiplicative smearing factor, one can exploit
the fact that the smearing function $f$ is sharply peaked about $y=1$
to formally write
\begin{equation}
f(y,\gamma)\ =\ {\cal N}\ \delta(y-1)\ +\ \delta f(y,\gamma)\ ,
\end{equation}
where ${\cal N} = \int_0^{M_A/M} dy\ f(y,\gamma)$ is the normalization
of the smearing function, which for $\gamma=1$ is either unity for the
unpolarized $F_2$ structure function, or equal to the effective nucleon
polarization in the nucleus for the spin-dependent $g_1$ structure
function.
The correction $\delta f$ gives the finite width of the smearing function. 
The smeared neutron structure function in Eq.~(\ref{eq:Nconv}) can
then be written
\begin{equation}
\widetilde{{\cal F}}^n(x)\
=\ {\cal N}\ {\cal F}^n(x)\
+\ \left(\delta f \otimes {\cal F}^n\right)(x)\ .
\label{eq:Nconv:2}
\end{equation}
The convolution term in Eq.~(\ref{eq:Nconv:2}) can thus be treated
as a perturbation and the equation solved iteratively.
Starting from a first guess ${\cal F}^{n(0)}(x)$ one has, after one 
iteration,
\begin{equation}
{\cal F}^{n(1)}(x)\
=\ {\cal F}^{n(0)}(x)\
+\ \frac{1}{\cal N}
   \left[ \Fntwiddle - \left( f \otimes {\cal F}^{n(0)} \right)(x)
   \right]\ .
\label{eq:add}
\end{equation}
Here, there is no danger of divergences due to zeros in the input,
as the only division is by ${\cal N}$, which is nonzero for all
smearing functions $f(y,\gamma)$.

When ${\cal F} = xg_1$, Eq.~(\ref{eq:Nconv}) is a system of two equations,
whose solution is slightly more involved.
One notes that the function $f_{11}$ is the most sharply peaked of the
smearing functions \cite{KM_d}, and hence gives the largest contribution
to $xg_1^d$.
Assuming that the $f_{12}$ contribution is zero, one can apply
Eq.~(\ref{eq:add}) to $xg_1^d$ to obtain $g_1^{n(1)}(x)$, which is
substituted into the expression for $xg_2^d$.
Subtracting this contribution $f_{21} \otimes xg_1^{n(1)}$ from $xg_2^d$
and applying Eq.~(\ref{eq:add}) to the resulting expression then gives
$g_2^{n(1)}(x)$.
The new value $g_2^{n(1)}(x)$ is then inserted into the $xg_1^d$ equation
and the recursive procedure repeated until convergence is achieved.

\subsection{Analysis of convergence}
\label{ssec:converge}

As we will show in Sec.~\ref{sec:results} below, the convergence of
the additive method is quite fast and nearly independent of the 
initial guess.
The reason for this is essentially the sharply peaked shape of the
smearing function.
This can be illustrated by examining the propagation of the error on
the true function ${\cal F}_{\rm true}(x)$ with each iteration $i$.
Starting from a first guess, ${\cal F}^{(0)}(x)$, for the true
function, we define
${\cal F}^{(0)}(x) = {\cal F}_{\rm true}(x) + \epsilon^{(0)}(x)$,
where $\epsilon^{(0)}(x)$ is the difference between the first guess
and the true result.
Tracking this error after $i=1$ iteration gives
\begin{equation}
\epsilon^{(1)}(x)
= \epsilon^{(0)}(x)
- \frac{1}{\cal N}\ (f \otimes \epsilon^{(0)})(x)\ .
\label{eq:eps_add}
\end{equation}
Note that if $f(y) \sim \delta(y-1)$, the error for $x < 1$ vanishes 
even after one iteration.
In fact, since $f(y)$ is sharply peaked at $y = 1$ (for the unpolarized
and diagonal polarized distributions), $\epsilon^{(1)}(x)$ is expected
to be quite small for $x \lesssim 0.8$, regardless of ${\cal F}^{(0)}$.

More specifically, the iteration procedure will converge if for
successive iterations $|\epsilon^{(i+1)}(x)| < |\epsilon^{(i)}(x)|$.
Defining $\sigma$ to be the width over which the smearing function
$f(y) \gg 0$, from Eq.(\ref{eq:f0}) it follows that
$\sigma \sim \gamma p_{\rm char}/M \ll 1$, where $p_{\rm char}$
is the characteristic nucleon momentum scale in the deuteron.
Then using the generalized mean value theorem for integrals, the
correction term in Eq.~(\ref{eq:eps_add}) can be written as
\begin{equation}
{1 \over {\cal N}}\ (f \otimes \epsilon)(x)\
=\ \epsilon(x/y_*)\ ,
\label{eq:gmvt}
\end{equation}
where $y_* = 1+c$ is a point within the integration interval,
with $|c| < \sigma/2$.
If $\epsilon(x)$ is a sufficiently smooth function of $x$, one can
expand the right-hand-side of Eq.~(\ref{eq:gmvt}) in a series in $c$,
\begin{equation}
\epsilon(x/(1+c))\
=\ \epsilon(x) - c x\ \epsilon'(x) + {\cal O}(c^2)\ ,
\end{equation}
so that the error after one iteration is 
$\epsilon^{(1)}(x) \approx c x \epsilon^{(0)'}(x)$.
This then leads to the estimate 
\begin{equation}
\left| {\epsilon^{(1)} \over \epsilon^{(0)}} \right|\
\approx\ c x \left| {\epsilon^{(0)'} \over \epsilon^{(0)}} \right|\
 < \ {\sigma \over 2}
     \left| {\epsilon^{(0)'} \over \epsilon^{(0)}} \right|\ .
\end{equation}
so that the ratio of errors $\epsilon^{(1)}/\epsilon^{(0)}$ is
proportional to the width of the smearing function, as long as the
width is small.
Furthermore, because the $i=1$ error is given by the derivative of
$\epsilon^{(0)}$, convergence is fastest when the error is smoothest,
which will typically be away from resonance peaks.

%
%

\section{Results}
\label{sec:results}

In this section we present numerical results which illustrate the
features of the extraction methods discussed in Sec.~\ref{sec:methods}.
We discuss firstly the unpolarized $F_2^n$ structure function, before
considering the more challenging case of the polarized $g_1^n$
structure function.

\subsection{Unpolarized structure functions}
\label{ssec:F2}

\begin{figure}[t] 
\includegraphics[width=13cm]{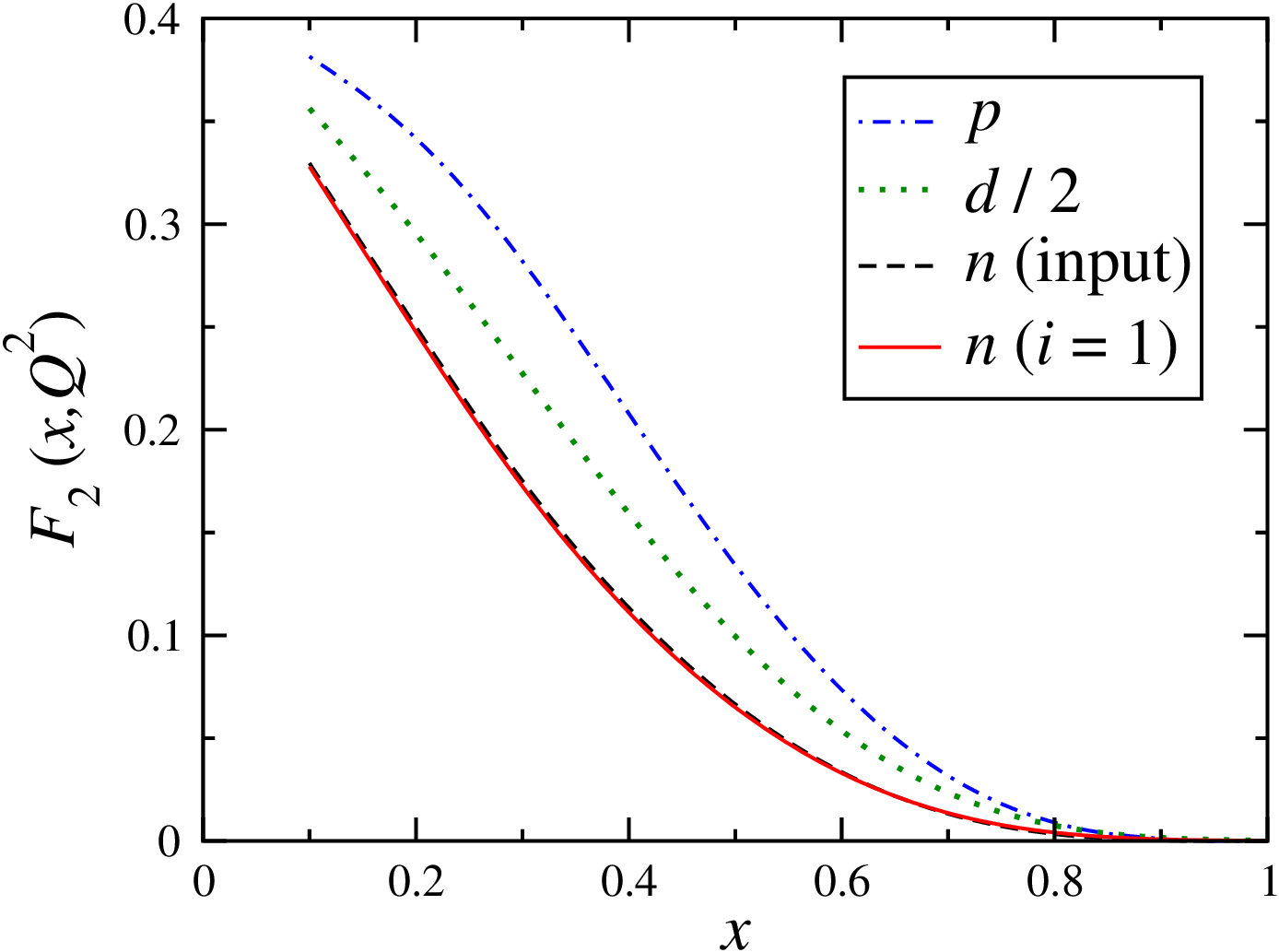}
\caption{(Color online)
    Extraction of the neutron $F_2^n$ structure function at
    $Q^2 = 10$~GeV$^2$ from $F_2^p$ (dashed) and $F_2^d$ (dot-dashed)
    data simulated from the MRST parameterization \cite{MRST} and the
    smearing function $f_0(\gamma,y)$ \cite{KP} using the additive
    method.  The extracted $F_2^n$ structure function after $i=1$
    iteration (solid) is almost indistinguishable from the input
    (dotted).}
\label{fig:F2dis}
\end{figure}

Most previous extractions of the $F_2^n$ structure function have been
performed in the deep inelastic region, where the structure functions
are smooth and monotonic (beyond $x \sim 0.3$).
Before applying our extraction procedure to the more challenging
resonance region, we first test the method on the more familiar case
of DIS kinematics.
For the input proton and neutron structure functions we use the MRST
parameterization \cite{MRST} at $Q^2 = 10$~GeV$^2$, and simulate the
deuteron $F_2^d$ ``data'' using the finite-$Q^2$ smearing function
$f_0(y,\gamma)$ from Ref.~\cite{KP}.

The resulting extracted neutron $F_2^n$ structure function is shown
in Fig.~\ref{fig:F2dis} using the additive method.
Starting from an initial guess of $F_2^{n(0)}=0$, the extracted curve
is almost indistinguishable from the input $F_2^n$ after just a single
iteration.
The main reason for this fast convergence is the fact that the nucleons
in the deuteron are weakly bound and have small average momentum, which
leads to a smearing function $f_0(y,\gamma)$ that is sharply peaked
around $y=1$.
Although the precise height and width of the peak may vary slightly for
different deuteron wave functions, the rapid convergence is a relatively
model-independent feature of the extraction.

\begin{figure}[t] 
\includegraphics[width=13cm]{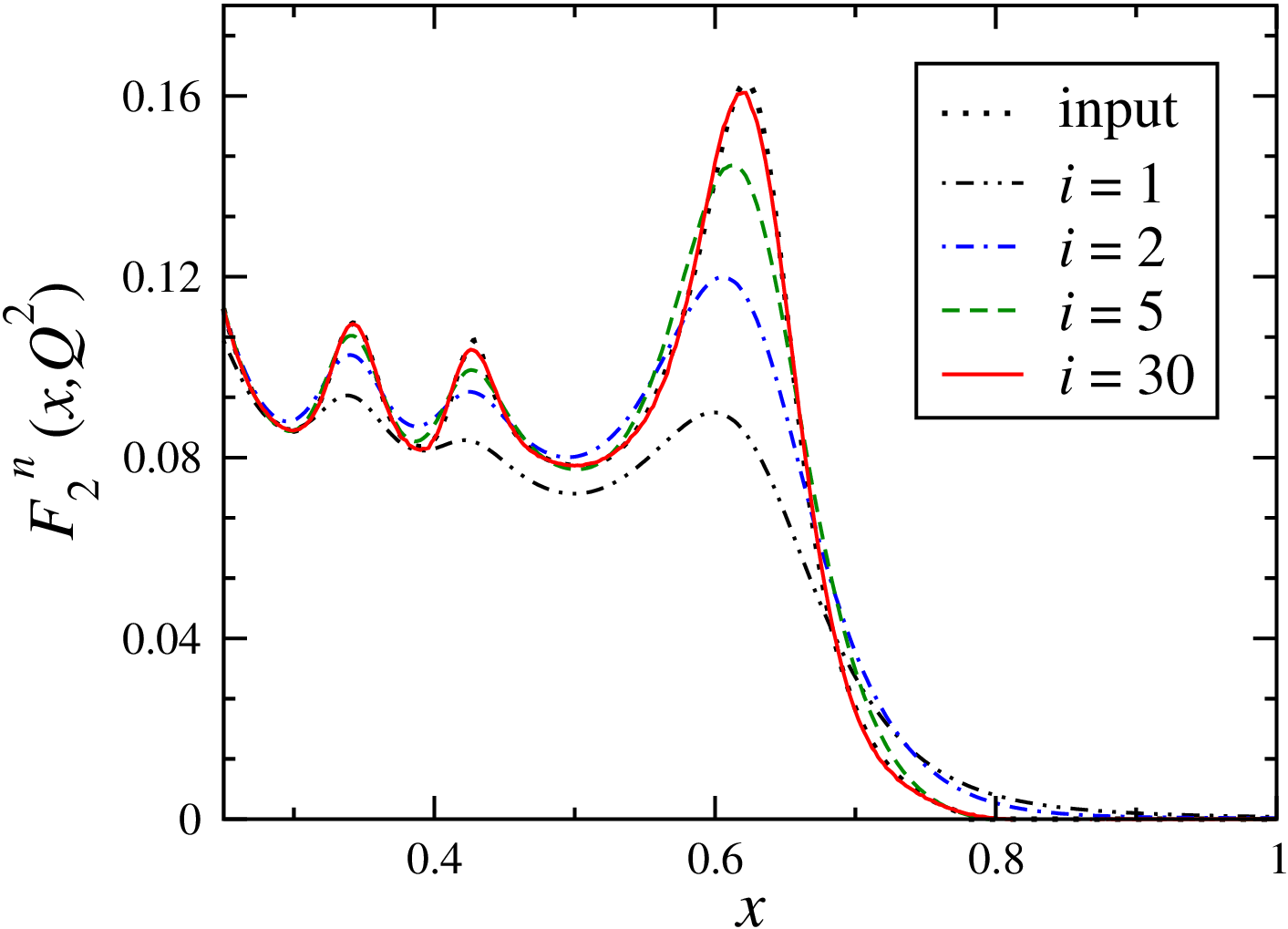}
\caption{(Color online)
    Extraction of the neutron $F_2^n$ structure function at
    $Q^2 = 1$~GeV$^2$ in the resonance region using the MAID
    parameterization \cite{MAID} for the input (dotted),
    and a first guess $F_2^{n(0)}=0$.
    The convergence of the procedure is illustrated by the
    results for $i=1$ (dot-dot-dashed), 2 (dot-dashed),
    5 (dashed) and 30 (solid) iterations.}
\label{fig:F2itn}
\end{figure}

While the extraction of $F_2^n$ in the deep inelastic region is
straightforward, obtaining $F_2^n$ in the nucleon resonance region,
where the cross section is dominated by resonance peaks, is more
problematic.
In fact, to our knowledge such an extraction has not yet been
undertaken in any quantitative analysis.
Even in a system as dilute as the deuteron, the structure of nucleon
resonances is significantly smeared out by the Fermi motion of the
nucleons, so that for $Q^2 \sim 1$~GeV$^2$ or higher essentially only
the $\Delta$ region exhibits any clear resonance structure.
In heavier nuclei there is very little resonance structure evident
at all \cite{F2A_EMC}.
It is not clear {\em a priori} therefore to what extent neutron
resonance data can be extracted from data in which the neutron
information is strongly smeared.

To test the effectiveness of the additive extraction method in the
resonance region we use as input structure functions from the MAID
Unitary Isobar Model \cite{MAID}, which is constructed to parameterize
meson electroproduction data at low $W$.
The convergence of the iteration procedure in the resonance region is
illustrated in Fig.~\ref{fig:F2itn}, where we attempt to extract the
input $F_2^n$ at $Q^2 = 1$~GeV$^2$ with an increasing number of
iterations.
Taking as a first guess $F_2^{n(0)}=0$, after $i=1$ or 2 iterations
the prominent resonant structures are clearly visible, although the
amplitudes of the resonance peaks is still underestimated.
After $i=5$ iterations the extracted function is very close to the true
result, and would in most cases lie within experimental uncertainties.
Repeating the procedure $i=30$ times reproduces the complete resonance
structures almost exactly.

\begin{figure}[t] 
\includegraphics[width=13cm]{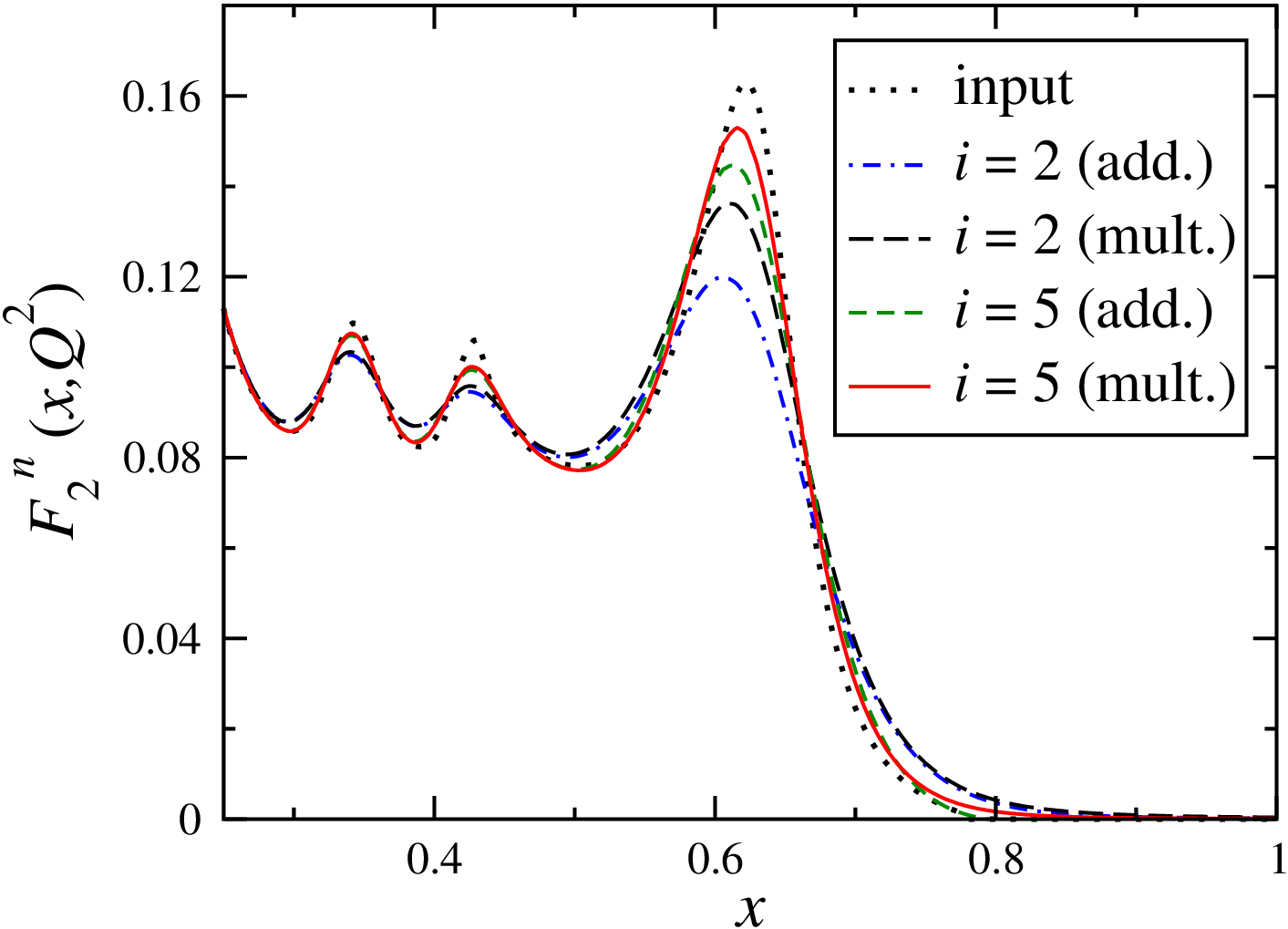}
\caption{(Color online)
    Comparison of the convergence of the additive and multiplicative
    methods for extraction $F_2^n$ for $i=2$ and 5 iterations.
    The starting point for the additive method was $F_2^{n(0)}=0$,
    while for the multiplicative method $F_2^{n(0)}=F_2^p$.
    The input structure functions were taken from the MAID
    parameterization \cite{MAID} at $Q^2 = 1$~GeV$^2$.}
\label{fig:F2addmult}
\end{figure}

The multiplicative method can also be used to extract $F_2^n$ in
the resonance region, as illustrated in Fig.~\ref{fig:F2addmult}.
The starting point for the iteration here is taken to be
$F_2^{n(0)}=F_2^p$, and after $i=5$ iterations the result is in good
agreement with the input function, only slightly underestimating
the peaks of the resonances.
As in the additive method, almost perfect agreement can be achieved
eventually with further iterations.
Note that a direct comparison of the convergence of the additive and
multiplicative methods from Fig.~\ref{fig:F2addmult} is not possible
since the starting points $F_2^{n(0)}$ are different.
Here we merely illustrate the fact that both methods can converge
to the true result within a relatively small number of iterations.

\begin{figure}[t] 
\includegraphics[width=13cm]{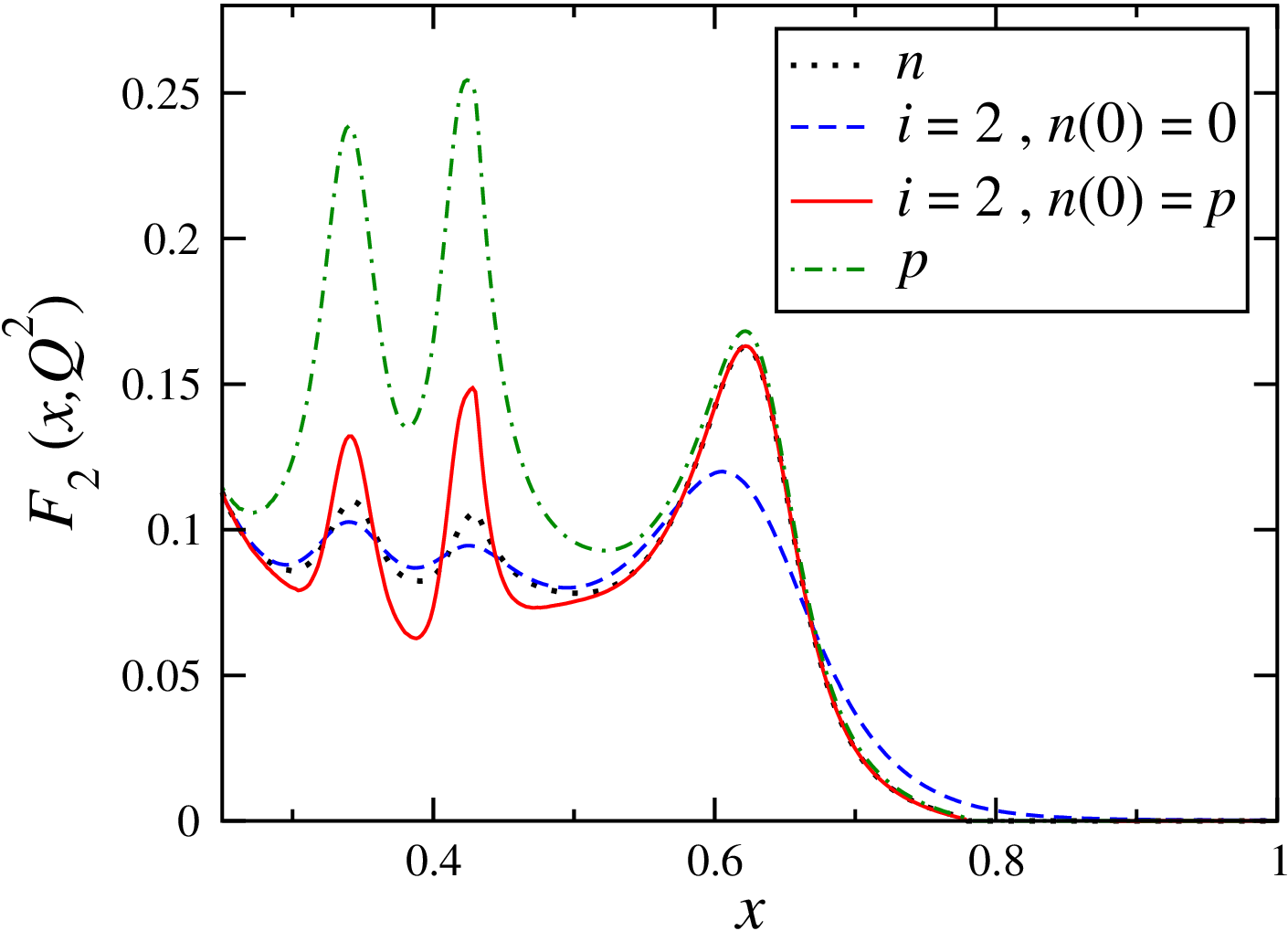}
\caption{(Color online)
    Extracted neutron $F_2^n$ structure function using the
    additive method with $i=2$ iterations, starting with initial
    guesses $F_2^{n(0)}=0$ (labeled ``$n(0)=0$'', dashed) and
    $F_2^{n(0)}=F_2^p$ (labeled ``$n(0)=p$'', solid).
    The input neutron (dotted) and proton (dot-dashed) structure
    functions are taken from the MAID parameterization \cite{MAID}
    at $Q^2 = 1$~GeV$^2$.}
\label{fig:F2guess}
\end{figure}

To examine the sensitivity of the extraction to the initial guess
$F_2^{n(0)}$, in Fig.~\ref{fig:F2guess} we show the result after $i=2$
iterations for initial guesses $F_2^{n(0)}=0$ and $F_2^{n(0)}=F_2^p$,
using the MAID fit \cite{MAID} at $Q^2=1$~GeV$^2$ as input.
Since the amplitudes of the resonances are significantly larger for
the proton than for the neutron, the $F_2^p$ initial guess results
in larger amplitudes for the extracted neutron $F_2^n$ for the same
number of iterations.
On the other hand, because the proton and neutron resonance transitions
to the $\Delta$ are expected to be equal (since the transitions are
isovector), the proton initial guess enables the $\Delta$ peak to be
reproduced extremely well, in contrast to the zero first guess which
requires more iterations to produce the observed structure.
Of course, with sufficiently many iterations the input $F_2^n$ can be
reproduced accurately regardless of the initial guess $F_2^{n(0)}$.

\begin{figure}[t] 
\includegraphics[width=13cm]{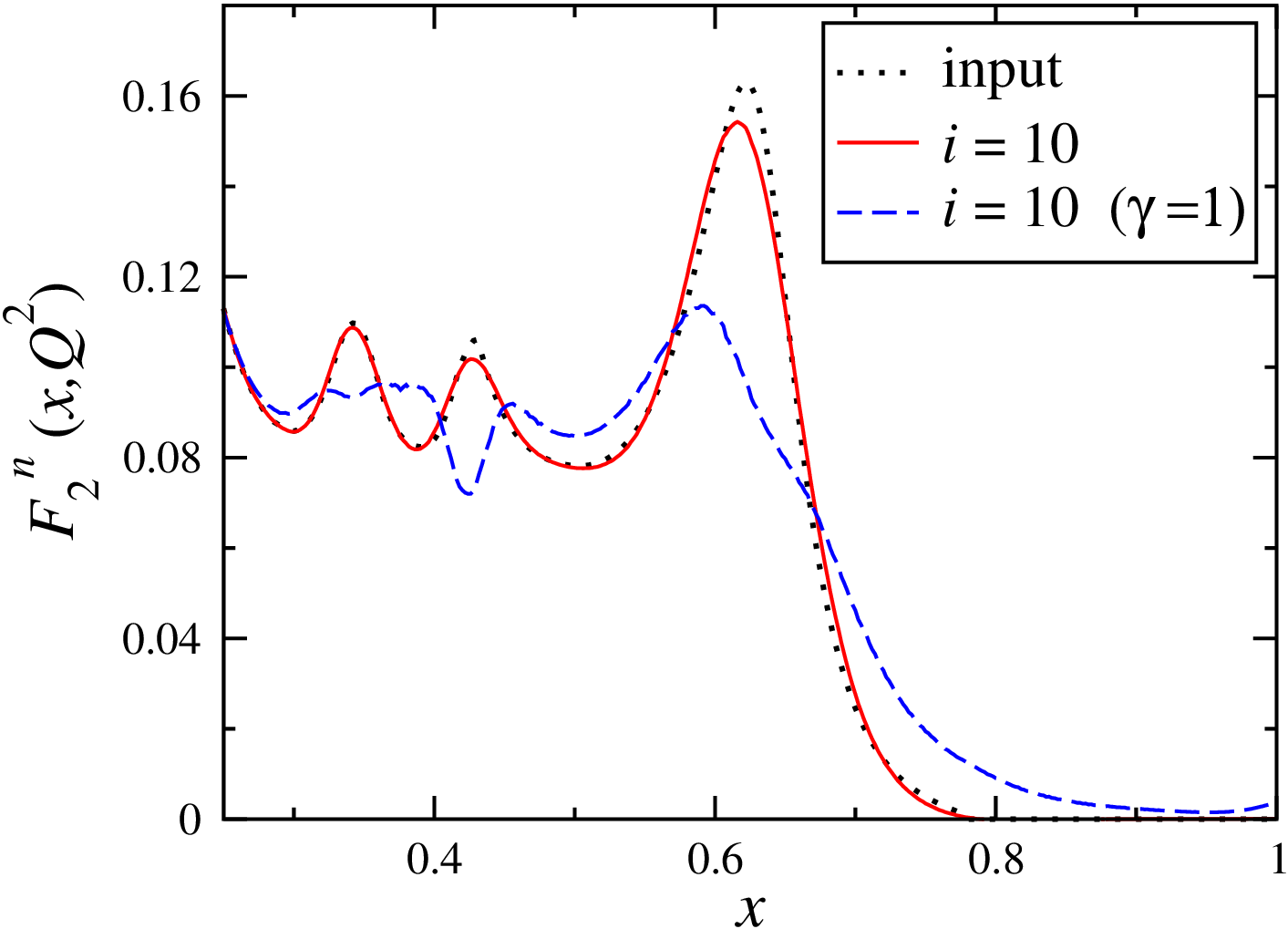}
\caption{(Color online)
    Extracted neutron $F_2^n$ structure function using the
    additive method after $i=10$ iterations with the full
    $\gamma$-dependent smearing function (solid) and with
    the $\gamma=1$ approximation (dashed), compared with the
    input neutron (dotted) structure function from the MAID
    parameterization \cite{MAID} at $Q^2 = 1$~GeV$^2$.}
\label{fig:F2gam}
\end{figure}

In all of the above extractions the full $\gamma$- (or $Q^2$-) dependent
nucleon smearing function $f_0(y,\gamma)$ has been used when computing
the deuteron structure function.
While using a $\gamma$-independent smearing function may be a reasonable
approximation in the deep inelastic region where $\gamma$ values are
typically close to unity, applying the $\gamma=1$ smearing function to
low-$Q^2$, large-$x$ data can lead to errors in the extracted $F_2^n$,
especially in the resonance region \cite{Bosted}.

The importance of using the correct smearing function is illustrated
in Fig.~\ref{fig:F2gam}, where we show the extracted neutron $F_2^n$
structure function after $i=10$ iterations.
The result using the full, $\gamma$-dependent smearing function is
very close to the input.
On the other hand, with the $Q^2$-independent, $\gamma=1$ smearing
function the iteration does not converge to the correct solution.
In particular, while a resonance bump is visible in the $\Delta$ region,
it has the incorrect strength; the second resonance region displays
a trough where there should be a peak; and the third resonance region
appears to have no structure at all.
Increasing the number of iterations for the $\gamma$-dependent
smearing function leads to ever closer convergence to the input $F_2^n$.
For the $\gamma=1$ smearing function, the result does not change
qualitatively with further iterations, however significant noise
develops over much of the $x$ range.

These features arise from the mismatch between the smearing functions
used to compute the deuteron $F_2^d$ and those used to perform the
extraction.
Of course, had the deuteron structure function been simulated with the
$\gamma=1$ smearing function, the extraction with the same function
would return the same input $F_2^n$ as in Fig.~\ref{fig:F2gam}.
However, this comparison demonstrates the sensitivity of the extraction
to the $Q^2$ dependence of the smearing function, and highlights the
importance of using a smearing function with the correct $Q^2$
dependence when analyzing actual data \cite{F2data}.

\subsection{Polarized structure functions}
\label{ssec:g1}

\begin{figure}[t] 
\includegraphics[width=13cm]{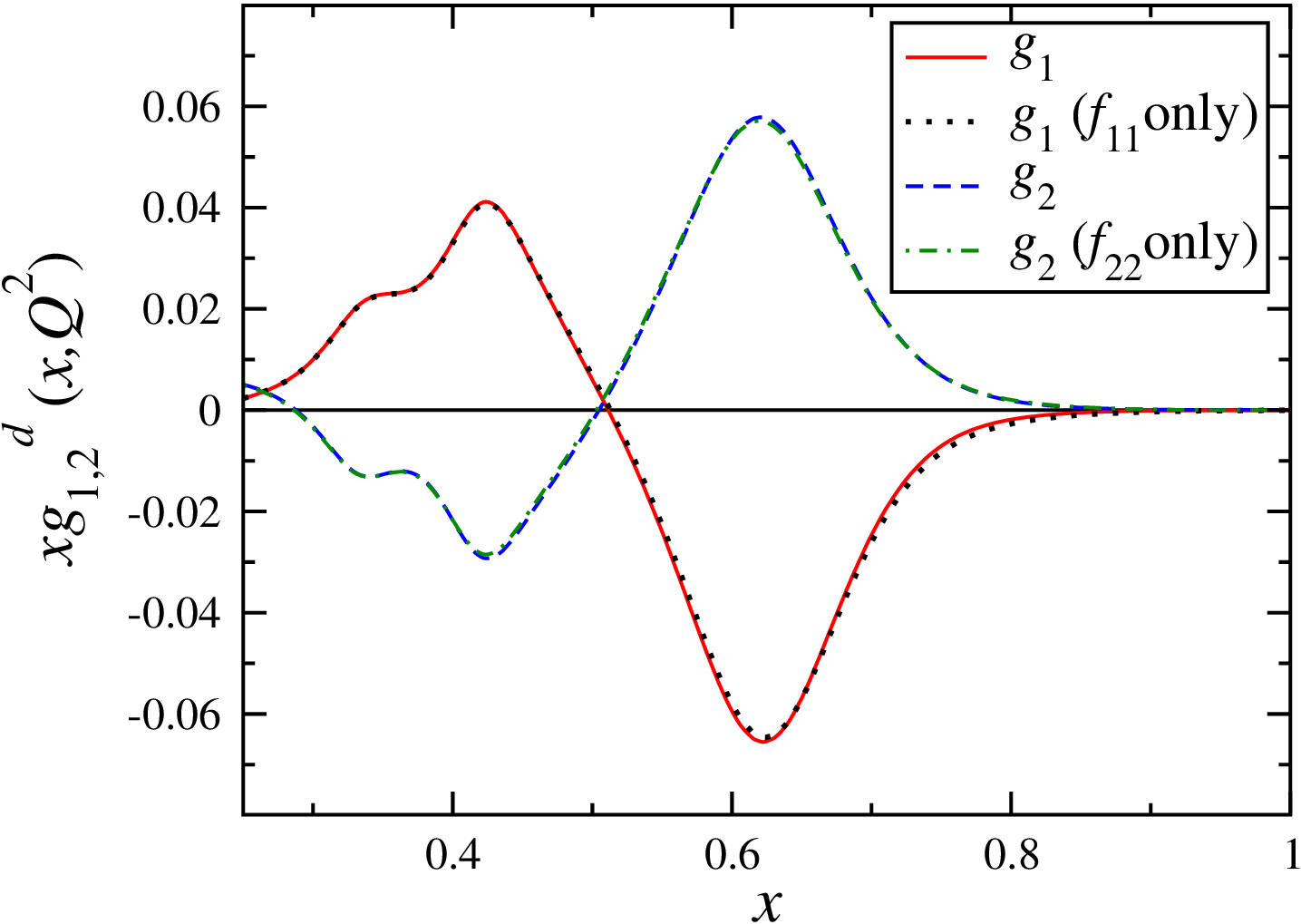}
\caption{(Color online)
    Deuteron $xg_1^d$ and $xg_2^d$ structure functions simulated
    from the MAID parameterization of the proton and neutron
    $g_{1,2}$ at $Q^2=1$~GeV$^2$ and the smearing functions
    $f_{ij}$, $i,j=1,2$, from Ref.~\cite{KM_d}.
    The full results for $xg_1^d$ (solid) and $xg_2^d$ (dashed)
    are compared with the diagonal approximations using $f_{11}$
    only (dotted) and $f_{22}$ only (dot-dashed), respectively.}
\label{fig:g12d}
\end{figure}

At finite $Q^2$ the complete expressions for the $g_1^d$ and $g_2^d$
nuclear structure functions in Eq.~(\ref{eq:conv_g}) represent a
coupled set of equations involving contributions from both the $g_1^N$
and $g_2^N$ structure functions of the nucleon \cite{KM_d}.
While the diagonal $f_{11}$ and $f_{22}$ smearing functions dominate for
most kinematics, the off-diagonal $f_{12}$ and $f_{21}$ contributions
could be important at low values of $Q^2$.
Furthermore, the $g_1^N$ contribution to $g_2^d$ survives even in the
Bjorken limit.

As described in Sec.~\ref{ssec:add} above, one can solve such a system
of equations by simultaneously iterating both $g_1^n$ and $g_2^n$, given
known (or simulated) proton and deuteron data.
Such a procedure will necessarily be slower and require more iterations,
but is stable and will in principle converge to the correct solutions.

In practice, however, for the kinematics discussed here, namely
$Q^2 \sim 1$--10~GeV$^2$, the off-diagonal contributions are rather small.
This can be seen in Fig.~\ref{fig:g12d} where we show the $xg_1^d$ and
$xg_2^d$ structure functions simulated from the MAID $g_{1,2}^{p,n}$
parameterizations \cite{MAID} at $Q^2 = 1$~GeV$^2$, using the smearing
functions $f_{ij}(y,\gamma)$, $i,j=1,2$, from Ref.~\cite{KM_d}.
The results with the diagonal terms only ($f_{11}$ for $g_1^d$ and
$f_{22}$ for $g_2^d$), are very close to the full results which include
both diagonal and off-diagonal contributions.
With the precision achievable in current and near-term future experiments,
the diagonal approximation to the $g_{1,2}^d$ structure functions should
therefore provide a reliable framework in which to extract neutron
structure functions, and in the following analysis we consider only the
diagonal contributions.
Furthermore, since the shape of $g_2$ is qualitatively similar to that
of $g_1$ (generally $g_2$ has the opposite sign compared with $g_1$),
we shall focus on the $g_1$ structure function as representative of the
effects of extracting spin-dependent neutron structure functions in the
resonance region.

As we saw in the previous section, both the additive and multiplicative
methods yield reliable results for extracted neutron structure functions,
in both the deep inelastic and resonance regions, as long as the structure
functions are free of zeros.
For polarized scattering the $g_1$ and $g_2$ structure functions are no
longer positive-definite, so that taking ratios of smeared to unsmeared
functions can in principle lead to singularities during the extraction.

This does not necessarily render the multiplicative method completely
impractical for extracting polarized structure functions, however.
Numerically, for a given iteration where the structure function is
close to (but not exactly at) its zero, the smearing factor $S^n$ will
be very large.
For the next step in the iteration this large contribution will be
damped by the corresponding small value of the structure function,
making the result finite.
On the other hand, precisely how (and whether) this cancellation occurs
in practice will be determined by the shapes of the input structure
functions and smearing functions, and {\em a priori} it is not clear
whether an extracted non-positive definite structure function will
be well-behaved for a particular extraction.

\begin{figure}[t] 
\includegraphics[width=13cm]{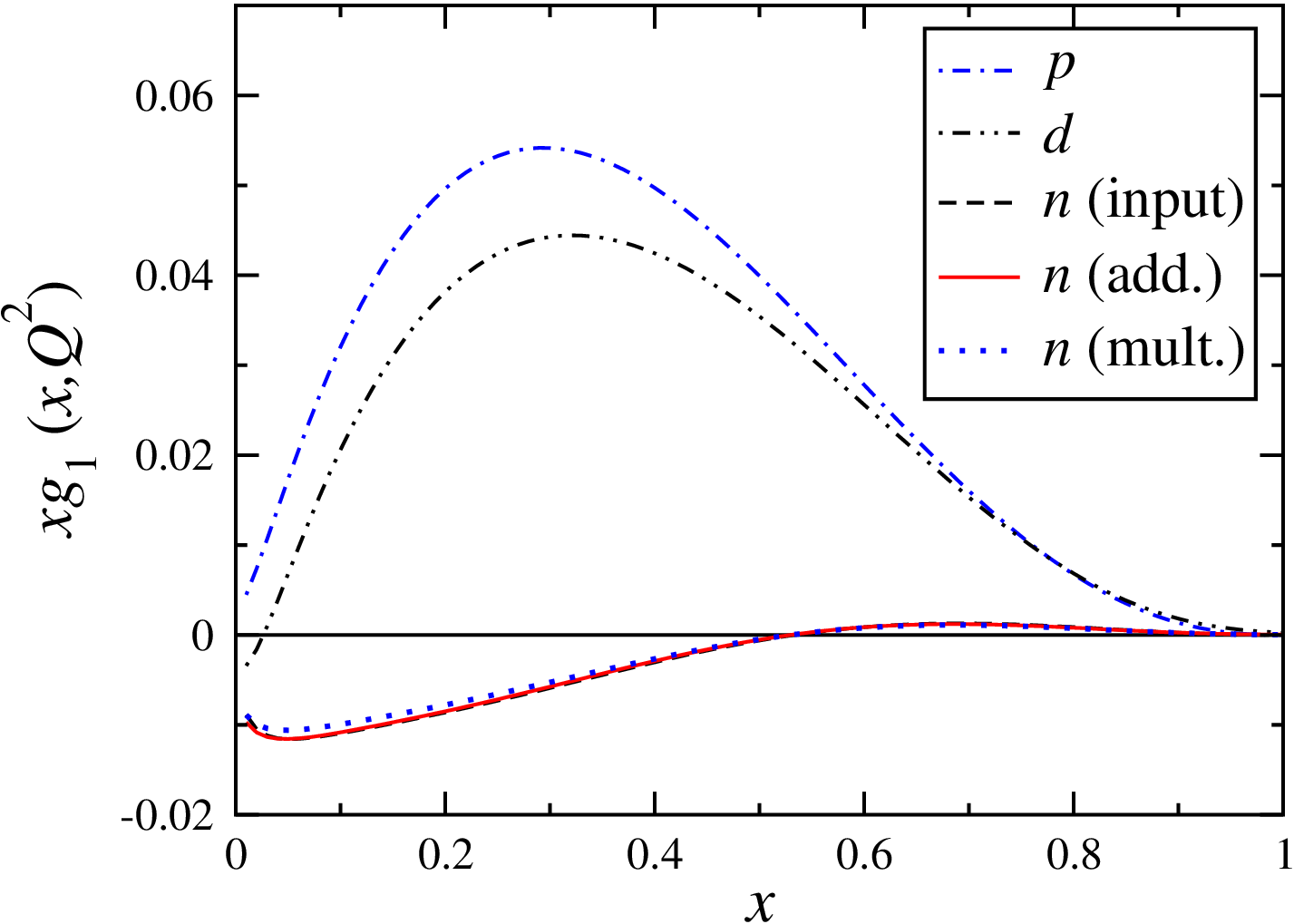}
\caption{(Color online)
    Extraction of the neutron $xg_1^n$ structure function
    from proton $xg_1^p$ (dot-dashed) and deuteron $xg_1^d$
    (dot-dot-dashed) ``data'' simulated from the leading twist
    parameterization \cite{BB} at $Q^2 = 10$~GeV$^2$ and the
    smearing function $f_{11}(\gamma,y)$ \cite{KP}.
    The input (dotted) and extracted $xg_1^n$ functions, for both
    the additive (solid) and multiplicative (dashed) methods, are
    almost indistinguishable.}
\label{fig:g1dis}
\end{figure}

To illustrate the extraction of spin-dependent structure functions
we first consider the $g_1$ structure function in the DIS region in
Fig.~\ref{fig:g1dis}.
The input proton $xg_1^p$ and neutron $xg_1^n$ data are taken from the
leading twist parameterization in Ref.~\cite{BB} at $Q^2 = 10$~GeV$^2$,
with the deuteron $xg_1^d$ simulated using the smearing function
$f_{11}(y,\gamma)$ from Ref.~\cite{KM_d}.
With a starting point of $xg_1^{n(0)} = 0$, the extracted neutron
structure function after a single iteration using the additive method
is essentially indistinguishable from the input.
As for the unpolarized $F_2$ structure function in the DIS region in
Fig.~\ref{fig:F2dis}, this feature reflects the narrow width of the
smearing function $f_{11}(y,\gamma)$ around $y=1$.

For the multiplicative method the initial guess is taken to be
$xg_1^{n(0)} = xg_1^p$, and after one iteration the extracted
neutron structure function is also very close to the input.
In particular, even though the ratio of smeared to unsmeared $g_1^n$
structure functions is singular at $x \approx 0.5$, the extracted
function is nevertheless continuous in this region.
The marginally slower convergence here compared with the additive
case reflects the different starting inputs for $g_1^n$, which for
the multiplicative method is further from the true result than for
the additive.

\begin{figure}[t] 
\includegraphics[width=13cm]{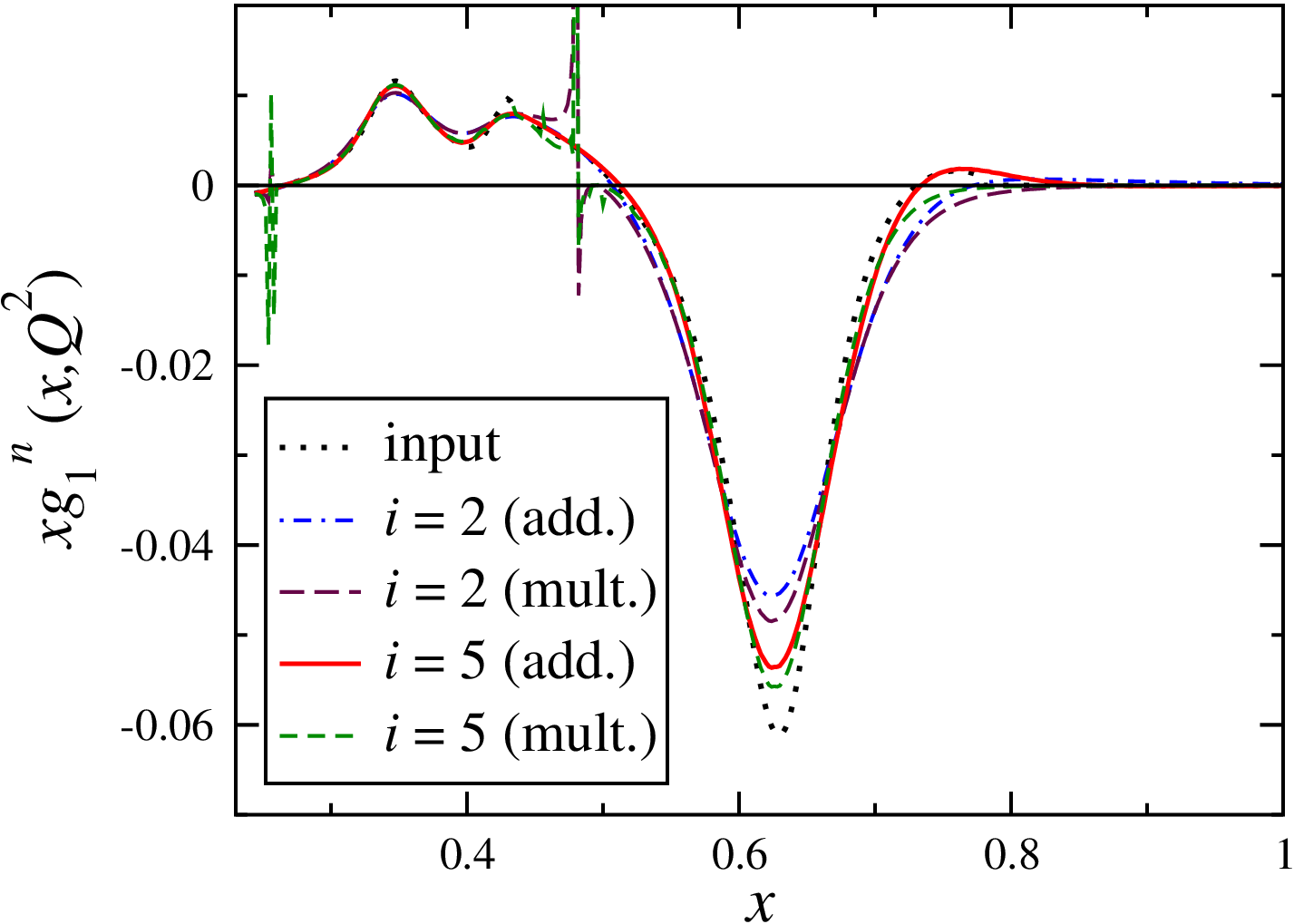}
\caption{(Color online)
    Comparison of the convergence of the additive and multiplicative
    methods for extraction $xg_1^n$ for $i=2$ and 5 iterations.
    The starting point for the additive method was $xg_1^{n(0)}=0$,
    while for the multiplicative method $xg_1^{n(0)}=xg_1^p$.
    The input structure functions were taken from the MAID fit
    \cite{MAID} at $Q^2 = 1$~GeV$^2$.}
\label{fig:g1addmult}
\end{figure}

While both the additive and multiplicative methods appear to be
effective in extracting the spin-dependent neutron structure function
in the DIS region, their utility in the nucleon resonance region,
where the $xg_1$ exhibits considerably more structure, is compared in
Fig.~\ref{fig:g1addmult} using the MAID parameterization \cite{MAID}
at $Q^2=1$~GeV$^2$.
The most striking feature of the extracted neutron $xg_1^n$ is
the discontinuities near the zeros of the input function for the
multiplicative method, which arise from the singularities in the
smearing factor $S^n$.
On the other hand, no such singularities appear for the additive
method and the extracted functions are smooth and continuous over
the entire range of $x$.

\begin{figure}[t] 
\includegraphics[width=13cm]{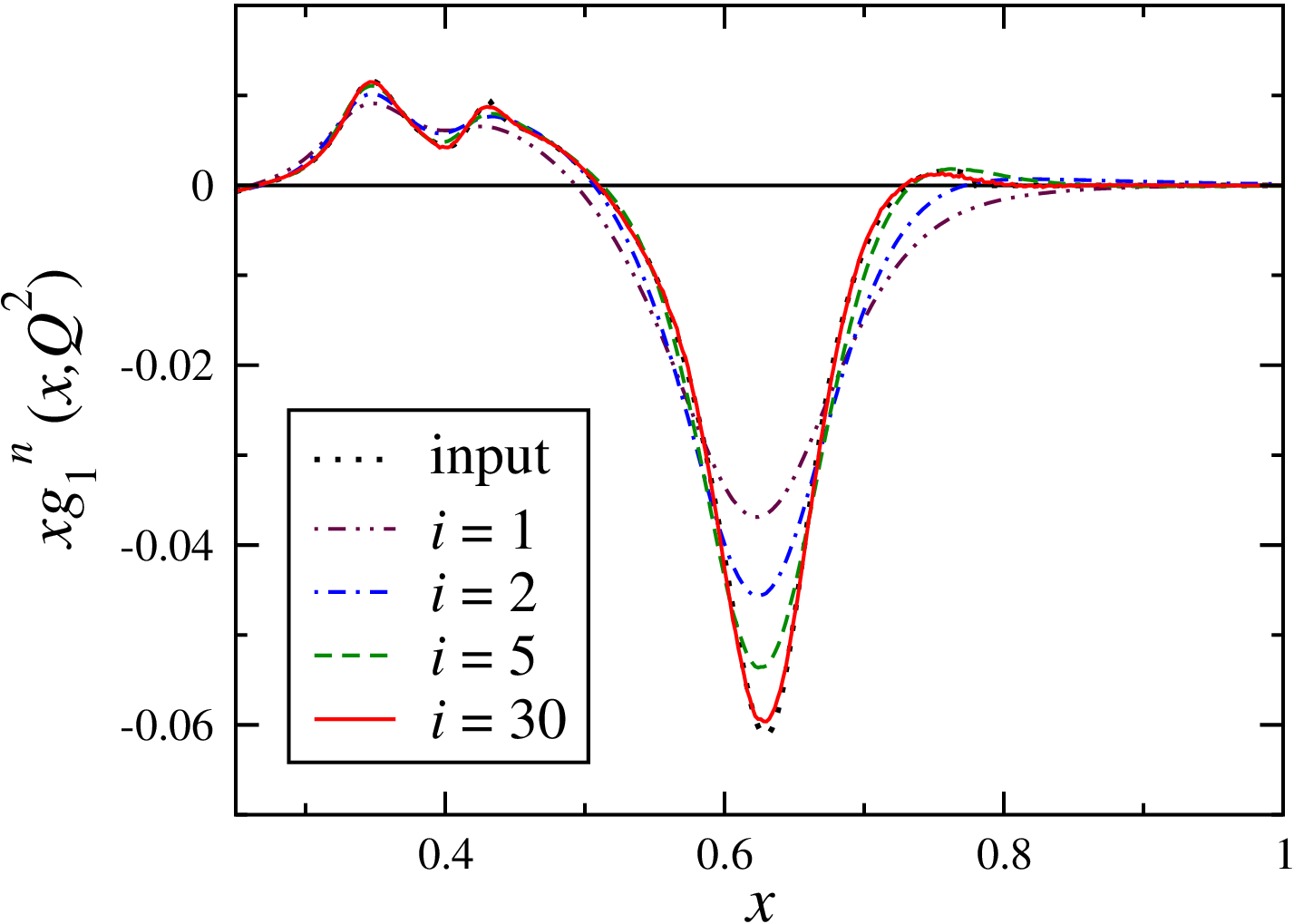}
\caption{(Color online)
    Convergence of the extracted neutron $xg_1^n$ structure
    function for $i=1$ (dot-dot-dashed), 2 (dot-dashed),
    5 (dashed) and 30 (solid) iterations, using the MAID
    resonance fit \cite{MAID} at $Q^2 = 1$~GeV$^2$ as input
    (dotted), with a first guess $xg_1^{n(0)} = 0$.}
\label{fig:g1itn}
\end{figure}

The convergence of the extraction for the additive method is illustrated
in Fig.~\ref{fig:g1itn}, where after only five iterations the extracted
$xg_1^n$ displays all of the prominent features of the $\Delta$ peak and
the higher resonance regions.
After $i=30$ iterations the input function is reproduced almost exactly.

\begin{figure}[t] 
\includegraphics[width=13cm]{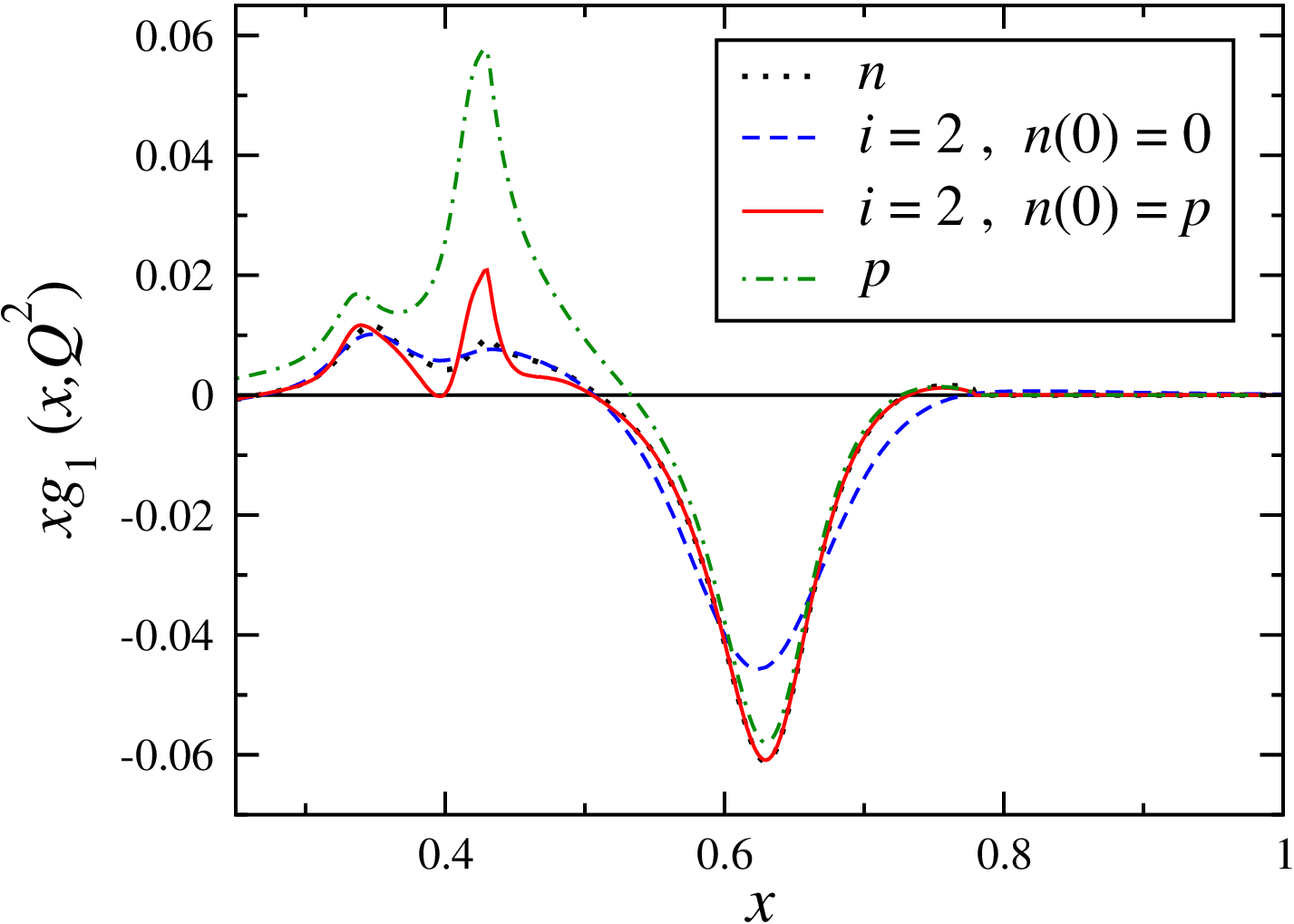}
\caption{(Color online)
    Extracted neutron $xg_1^n$ structure function using the
    additive method with $i=2$ iterations, starting with initial
    guesses $xg_1^{n(0)}=0$ (labeled ``$n(0)=0$'', dashed) and
    $xg_1^{n(0)}=xg_1^p$ (labeled ``$n(0)=p$'', solid).
    The input neutron (dotted) and proton (dot-dashed) structure
    functions are taken from the MAID fit \cite{MAID} at
    $Q^2 = 1$~GeV$^2$.}
\label{fig:g1guess}
\end{figure}

The sensitivity of the extraction to the initial guess for $xg_1^n$
is illustrated in Fig.~\ref{fig:g1guess}, where the results after
$i=2$ iterations are compared for starting values $xg_1^{n(0)}=0$
and $xg_1^{n(0)}=xg_1^p$.
As in the case of the unpolarized $F_2$ structure function, the
$xg_1^p$ initial guess gives amplitudes that are larger than for
the zero initial guess after the same number of iterations.
Since the input proton and neutron structure functions are similar
in the $\Delta$ region, the iteration of $xg_1^n$ converges on the
$\Delta$ peak more rapidly for the $xg_1^p$ starting point than for
the zero first guess.
On the other hand, because the second resonance peak for the proton
is significantly larger than for the neutron, convergence on this
is faster for the $xg_1^{n(0)}=0$ starting value.
Again, with sufficiently many iterations the input $xg_1^n$ can be
accurately reproduced independently of the starting point.

\begin{figure}[t] 
\includegraphics[width=13cm]{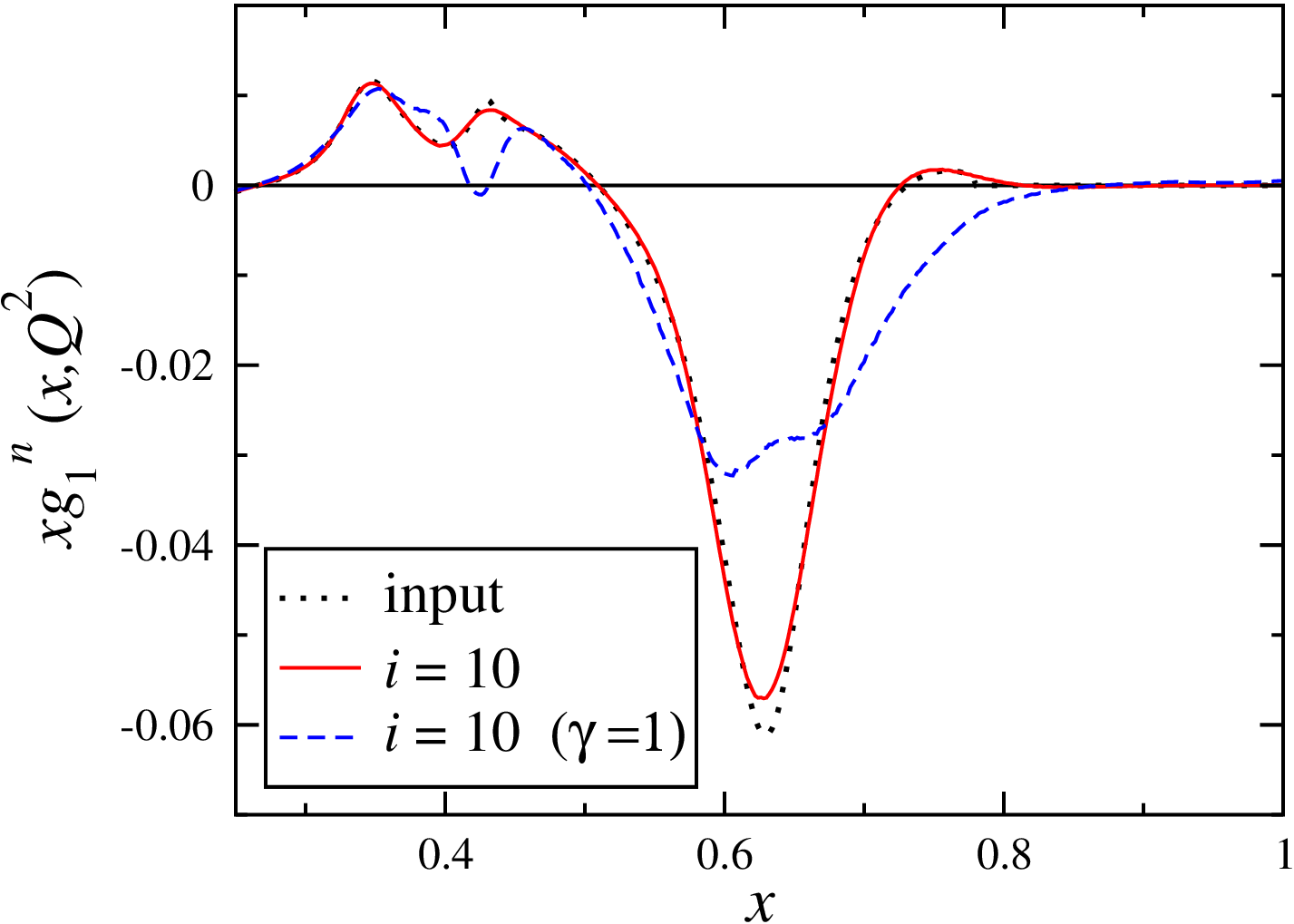}
\caption{(Color online)
    Extracted neutron $xg_1^n$ structure function using the
    additive method after $i=10$ iterations with the full
    $\gamma$-dependent smearing function (solid) and with
    the $\gamma=1$ approximation (dashed), compared with the
    input neutron (dotted) structure function from the MAID
    parameterization \cite{MAID} at $Q^2 = 1$~GeV$^2$.}
\label{fig:g1gam}
\end{figure}

The importance of using the correct $Q^2$ dependence in the smearing
function $f_{11}(y,\gamma)$ is highlighted in Fig.~\ref{fig:g1gam},
where the extracted $xg_1^n$ neutron structure function is shown after
$i=10$ iterations.
While the full, $\gamma$-dependent smearing function yields an almost
exact reconstruction of the input structure function, the result using
the $\gamma=1$ smearing function bears little resemblance to the true
$xg_1^n$.
Most notably, the height of the $\Delta$ peak is significantly
underestimated, and the position of the second resonance peak does not
correspond to the correct value.
As for the $F_2^n$ structure function in Fig.~\ref{fig:F2gam}, these
features arise from the mismatch between the smearing functions used to
compute the deuteron $xg_1^d$ and those used to perform the extraction
of $xg_1^n$.
They clearly demonstrate that it is vital to use the correct $Q^2$
dependence in the smearing function when analyzing data in the nucleon
resonance region, especially at low $Q^2$ and large $x$ \cite{g1d}.

\begin{figure}[t] 
\includegraphics[width=13cm]{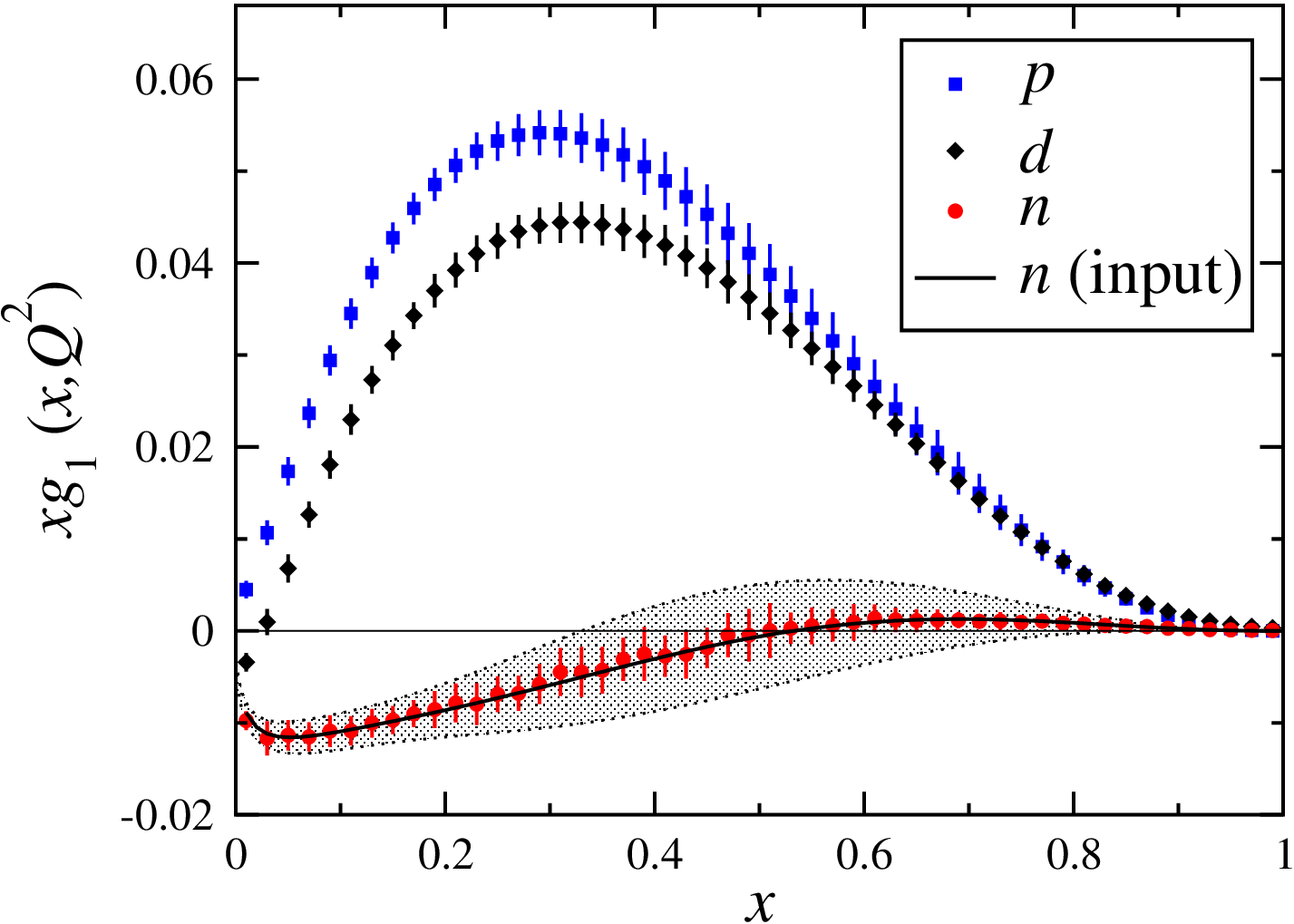}
\caption{(Color online)
    Extraction of the neutron $xg_1^n$ structure function (circles)
    from proton $xg_1^p$ (squares) and deuteron $xg_1^d$ (diamonds)
    ``data'' simulated from the leading twist parameterization
    \cite{BB} at $Q^2 = 10$~GeV$^2$ and the smearing function
    $f_{11}(\gamma,y)$ \cite{KP}.
    The error bars are derived from the uncertainties on the
    structure functions given in Ref.~\cite{BB}.
    The input $xg_1^n$ structure function (solid) is given as
    reference, with uncertainties indicated by the shaded band.}
\label{fig:g1data}
\end{figure}

Analysis of actual $g_1^d$ (and $F_2^d$) data to extract the free
neutron structure functions will be discussed in a forthcoming
publication \cite{F2data}.
However, one can anticipate how the neutron $xg_1^n$ structure function
can be extracted from actual proton and deuteron data, together with
error bars, by a simple illustration.

In Fig.~\ref{fig:g1data} we show the proton and deuteron structure
functions simulated from the leading twist parameterization \cite{BB}
at $Q^2 = 10$~GeV$^2$, with the error bars derived from the uncertainties
on the $xg_1^p$ and $xg_1^n$ structure functions given in Ref.~\cite{BB}.
The deuteron structure function was simulated by varying each point of
the proton and neutron input by a Gaussian of width given by the error
bar, which were then smeared with the momentum distribution $f_{11}$
and added to get a ``trial'' $xg_1^d$.
This procedure was repeated for 50 trials, after which the average and
standard deviation of each point was taken to obtain the $xg_1^d$ curves
and error bars.

To extract $xg_1^n$, one can assume that the only errors that contribute
are those from the deuteron.
In practice, errors on $xg_1^d$ are much larger than those on $xg_1^p$,
and smearing $xg_1^p$ renders the proton errors negligible compared to
the deuteron errors.
As before, each point of $xg_1^d$ was varied by the error bars, from
which the smeared $xg_1^p$ was then subtracted with no errors, and the
extraction performed to obtain a ``trial'' $xg_1^n$.
This was repeated for 50 trials and the average and standard deviation
computed as before.

The extracted neutron data points in Fig.~\ref{fig:g1data} are found to
be in excellent agreement with the input $xg_1^n$ structure function.
The errors on the extracted neutron function after one iteration are of
the same order of magnitude as those on the deuteron.
Note that the extracted error bars are considerably smaller than the
original error bars, indicated by the shaded band around the input
$xg_1^n$, which is mostly due to the fact that, as a sum of
smeared functions, the simulated $xg_1^d$ has artificially small errors.
For real data, errors on the deuteron and proton structure functions are
given, and neglecting the errors on $xg_1^p$ with respect to $xg_1^d$ is
a very reasonable assumption.

\section{Conclusions}
\label{sec:conclusion}

In this paper, we have presented a new method which allows the reliable
extraction of neutron structure functions, both spin-averaged and
spin-dependent, over a wide range of $Q^2$.
We have compared the new (additive) method to the existing
(multiplicative) extraction method, and found that the performance of
both methods is very similar for the extraction of $F_2^n$, while the
additive method is free of the singularities that develop when
attempting to extract $xg_1^n$ using the multiplicative method.
Moreover, the speed of convergence of the additive method is nearly
independent of the initial guess, and in most cases a reliable
extraction is achieved after $i=5$ iterations.
Finally, the extraction of $xg_1^n$ including error bars, shown in
Fig.~\ref{fig:g1data}, illustrates both that errors on the extracted
function can be reliably estimated, and that the performance of the
method is not overly sensitive to perturbations in the input.

The tests of the additive method on models of resonance-region structure
functions show that the general shape of the curve is reproduced after
only one iteration, but that further iterations are necessary to
accurately extract the magnitude of the resonance peaks.
The dependence of the method on the initial guess is evident in the
sense that the regions where convergence is slowest are the regions
where the initial guess is farthest from the actual magnitude of the
resonance peaks.
Thus, one can reduce the number of iterations needed with an educated
guess about the shape of the neutron function.
For example, since the isovector transition to the $\Delta$ gives
identical proton and neutron structure functions for the resonant part
of the $\Delta$, a good first guess for the neutron would always be the
proton structure function in the $\Delta$ region.
To ensure that the extracted neutron structure function is in fact
correct in the context of the smearing-function model, one can smear
the extracted structure function and add to the smeared proton structure
function to compare with the deuteron data.

Despite extensive experiments on light nuclear targets, the neutron
remains something of a mystery.
The same observables which can be directly measured for the proton must
be inferred for the neutron, because its instability outside of the
nucleus makes neutron targets impossible.
Previously, the low statistics and large errors from experiments
designed to measure neutron observables limited the accuracy of measured
neutron structure functions far more than using a simplified model of
the nucleus to perform the extraction.
The situation has changed with recent experiments at JLab, and now
accounting for nuclear corrections in neutron structure function
extraction procedures is essential to obtain an accurate representation
of the neutron structure functions, especially in the resonance region.
In particular, we have shown that ignoring finite-$Q^2$ corrections to
nuclear structure functions leads to an extracted neutron structure
function which may bear little resemblance to the true shape.
To assess quark-hadron duality for the neutron to the same extent that
it has been verified for the proton, detailed knowledge of all neutron
structure functions in all kinematic regimes is needed.
The method presented in this paper, when applied to the most recent JLab
data, will be a first step in that direction.

\begin{acknowledgments}

Y.~K. would like to thank the SULI program, funded by the DOE Office
of Science.  This work was supported by the DOE contract No.
DE-AC05-06OR23177, under which Jefferson Science Associates, LLC
operates Jefferson Lab.
S.~K. was partially supported by the Russian Foundation for Basic 
Research, grant 06-02-16659.

\end{acknowledgments}



\begin{thebibliography}{99}

\bibitem{BG}
E.~D.~Bloom and F.~J.~Gilman,
Phys.\ Rev.\ Lett.\  {\bf 25}, 1140 (1970).

\bibitem{MEK}
W.~Melnitchouk, R.~Ent and C.~Keppel,
Phys.\ Rept.\  {\bf 406}, 127 (2005)

\bibitem{Theory}
F.~E.~Close and N.~Isgur,
Phys.\ Lett.\  B {\bf 509}, 81 (2001);
%
F.~E.~Close and W.~Melnitchouk,
Phys.\ Rev.\  C {\bf 68}, 035210 (2003);
%
S.~J.~Brodsky,
arXiv:hep-ph/0006310.

\bibitem{Exp_p}
P.~E.~Bosted {\it et al.}, 
Phys.\ Rev.\  C {\bf 75}, 035203 (2007);
%
N.~Bianchi, A.~Fantoni and S.~Liuti,
Phys.\ Rev.\  D {\bf 69}, 014505 (2004);
%
A.~Airapetian {\it et al.}, 
Phys.\ Rev.\ Lett.\  {\bf 90}, 092002 (2003);
%
S.~Liuti, R.~Ent, C.~E.~Keppel and I.~Niculescu,
Phys.\ Rev.\ Lett.\  {\bf 89}, 162001 (2002);
%
I.~Niculescu {\it et al.},
Phys.\ Rev.\ Lett.\  {\bf 85}, 1182, 1186 (2000);
%
G.~Ricco, M.~Anghinolfi, M.~Ripani, S.~Simula and M.~Taiuti,
Phys.\ Rev.\  C {\bf 57}, 356 (1998).

\bibitem{Ales}
A.~Psaker, W.~Melnitchouk, M.~E.~Christy and C.~Keppel,
Phys.\ Rev.\  C {\bf 78}, 025206 (2008).

\bibitem{Osipenko}
M.~Osipenko, W.~Melnitchouk, S.~Simula, S.~Kulagin and G.~Ricco,
Nucl.\ Phys.\  A {\bf 766}, 142 (2006).

\bibitem{Exp_He3}
P.~Solvignon {\it et al.} [Jefferson Lab E01-012 Collaboration],
arXiv:0803.3845 [nucl-ex].

\bibitem{Bodek}
A.~Bodek {\it et al.},
Phys.\ Rev.\  D {\bf 20}, 1471 (1979);
%
A.~Bodek and J.~L.~Ritchie,
Phys.\ Rev.\  D {\bf 23}, 1070 (1981).

\bibitem{Whitlow}
L.~W.~Whitlow, E.~M.~Riordan, S.~Dasu, S.~Rock and A.~Bodek,
Phys.\ Lett.\  B {\bf 282}, 475 (1992).

\bibitem{Umnikov}
A.~Y.~Umnikov, F.~C.~Khanna and L.~P.~Kaptari,
Z.\ Phys.\  A {\bf 348}, 211 (1994).

\bibitem{Scopetta}
C.~Ciofi degli Atti, L.~P.~Kaptari, S.~Scopetta and A.~Y.~Umnikov,
Phys.\ Lett.\  B {\bf 376}, 309 (1996).

\bibitem{MST}
W.~Melnitchouk, A.~W.~Schreiber and A.~W.~Thomas,
Phys.\ Rev.\  D {\bf 49}, 1183 (1994);
%
Phys.\ Lett.\  B {\bf 335}, 11 (1994);
%
W.~Melnitchouk and A.~W.~Thomas,
Phys.\ Lett.\  B {\bf 377}, 11 (1996).

\bibitem{KPW}
S.~A.~Kulagin, G.~Piller and W.~Weise,
Phys.\ Rev.\  C {\bf 50}, 1154 (1994).

\bibitem{MPT}
W.~Melnitchouk, G.~Piller and A.~W.~Thomas,
Phys.\ Lett.\ B {\bf 346}, 165 (1995);
%
G.~Piller, W.~Melnitchouk and A.~W.~Thomas,
Phys.\ Rev.\ C {\bf 54}, 894 (1996).

\bibitem{KMPW}
S.~A.~Kulagin, W.~Melnitchouk, G.~Piller and W.~Weise,
Phys.\ Rev.\ C {\bf 52}, 932 (1995).

\bibitem{KP}
S.~A.~Kulagin and R.~Petti,
Nucl.\ Phys.\  A {\bf 765}, 126 (2006).

\bibitem{AKL}
S.~I.~Alekhin, S.~A.~Kulagin and S.~Liuti,
Phys.\ Rev.\  D {\bf 69}, 114009 (2004).

\bibitem{KM_d}
S.~A.~Kulagin and W.~Melnitchouk,
Phys.\ Rev.\  C {\bf 77}, 015210 (2008).

\bibitem{KM_He3}
S.~A.~Kulagin and W.~Melnitchouk,
Phys.\ Rev.\  C {\bf 78}, 065203 (2008).

\bibitem{Paris}
M.~Lacombe {\em et al.},
Phys.\ Rev.\  C {\bf 21}, 861 (1980).

\bibitem{MEC}
L.~P.~Kaptari, A.~I.~Titov, E.~L.~Bratkovskaya and A.~Y.~Umnikov,
Nucl.\ Phys.\  A {\bf 512}, 684 (1990);
%
W.~Melnitchouk and A.~W.~Thomas,
Phys.\ Rev.\ D {\bf 47}, 3783 (1993).

\bibitem{Golak}
J.~Golak {\em et al.},
Phys.\ Rept.\  {\bf 415}, 89 (2005).

\bibitem{HLM80}
Y.~Horikawa, F.~Lenz and N.~C.~Mukhopadhyay,
Phys.\ Rev.\  C {\bf 22}, 1680 (1980).

\bibitem{Volterra}
P.~Linz, \emph{Analytical and Numerical Methods for Volterra Equations},
Philadelphia: SIAM, 1985.

\bibitem{MRST}
A.~D.~Martin, R.~G.~Roberts, W.~J.~Stirling and R.~S.~Thorne,
Eur.\ Phys.\ J.\  C {\bf 28}, 455 (2003).

\bibitem{MAID}
D.~Drechsel, O.~Hanstein, S.~S.~Kamalov and L.~Tiator,
Nucl.\ Phys.\  A {\bf 645}, 145 (1999).

\bibitem{F2A_EMC}
J.~Arrington, R.~Ent, C.~E.~Keppel, J.~Mammei and I.~Niculescu,
Phys.\ Rev.\  C {\bf 73}, 035205 (2006).

\bibitem{Bosted}
P.~E.~Bosted and M.~E.~Christy,
Phys.\ Rev.\  C {\bf 77}, 065206 (2008).

\bibitem{F2data}
Y.~Kahn {\em et al.},
in preparation.

\bibitem{BB}
J.~Bl\"umlein and H.~B\"ottcher,
Nucl.\ Phys.\  B {\bf 636}, 225 (2002).

\bibitem{g1d}
K.~V.~Dharmawardane {\it et al.},
Phys.\ Lett.\  B {\bf 641}, 11 (2006);
%
Jefferson Lab Experiment E93-009,
G.~Dodge, S.~Kuhn and M.~Taiuti spokespersons,
and S.~Kuhn and N.~Guler, private communication;
%
Jefferson Lab Experiment E01-006,
O.~Rondon-Aramayo spokesperson.

\end{thebibliography}
\end{document}